\documentclass[11pt]{article}

\usepackage{scalerel}
\usepackage{amsthm,mathrsfs,latexsym,amsxtra,graphicx,appendix,epstopdf,feynmf,subfig,hyperref,setspace,fix-cm}
\usepackage[normalem]{ulem}
\usepackage{longtable}
\usepackage{makeidx}
\usepackage{fullpage}
\usepackage{amsmath}
\usepackage{amssymb}
\usepackage{setspace}
\usepackage{bbm}
\usepackage{dsfont}
\usepackage{graphics}
\usepackage{epsfig}
\usepackage[font=footnotesize,labelfont=bf,justification=centerlast,width=.94\textwidth]{caption}
\usepackage{cite}
 \usepackage{multirow}
\usepackage{array,booktabs}
\usepackage[cal=boondox]{mathalfa}
\usepackage{color}
\usepackage{amsfonts}
\usepackage{mathtools}
\usepackage[makeroom]{cancel}
\usepackage{tikz}
\usepackage{tensor}
\usepackage{simplewick}
\usepackage{frcursive}
\usetikzlibrary{decorations.markings}

\usepackage{hyperref}

\hypersetup{
    bookmarks=true,%
    colorlinks,%
    citecolor=blue,%
    filecolor=blue,%
    linkcolor=blue,%
    urlcolor=blue
}
\allowdisplaybreaks
\newcommand{\ba}{\begin{align}}
\newcommand{\ea}{\end{align}}

\newcommand{\dd}{\mathrm{d}}

\newcommand{\DD}{\mathrm{D}}

\newcommand{\be}{\begin{equation}}
\newcommand{\ee}{\end{equation}}
\newcommand{\bes}{\begin{equation*}}
\newcommand{\ees}{\end{equation*}}
\newcommand{\bea}{\begin{eqnarray}}
\newcommand{\eea}{\end{eqnarray}}
\newcommand{\beas}{\begin{eqnarray*}}
\newcommand{\eeas}{\end{eqnarray*}}

\newcommand{\bmat}{\begin{bmatrix}}
\newcommand{\emat}{\end{bmatrix}}

\def\le{\left}
\def\ri{\right}

\def\le{\left}
\def\ri{\right}

\newcommand{\ben}{\begin{enumerate}}
\newcommand{\een}{\end{enumerate}}

\begin{document}
\numberwithin{equation}{section}
{
\begin{titlepage}
\begin{center}

\hfill \\
\hfill \\
\vskip 0.75in

{\Large \bf Gravitational Anomalies in nAdS$_2$/nCFT$_1$}\\

\vskip 0.4in

{\large Alejandra Castro and Beatrix M\"uhlmann}\\

\vskip 0.3in

{\it Institute for Theoretical Physics Amsterdam and Delta Institute for Theoretical Physics, University of Amsterdam, Science Park 904, 1098 XH Amsterdam, The Netherlands} \vskip .5mm

\texttt{a.castro@uva.nl, b.muhlmann@uva.nl}

\end{center}

\vskip 0.35in

\begin{center} {\bf ABSTRACT } \end{center}

We revisit the holographic description of the near horizon geometry of the BTZ black hole in AdS$_3$ gravity, with a gravitational Chern-Simons term included.   After a dimensional reduction of the three dimensional theory, we use the framework of nAdS$_2$/nCFT$_1$ to describe the near horizon physics. This setup allows us to contrast the role of the gravitational and conformal anomaly inherited from AdS$_3$/CFT$_2$ in the symmetry breaking mechanism of nAdS$_2$/nCFT$_1$.
 Our results display how boundary conditions in the 3D spacetime, combined with the gravitational anomaly, affect the holographic description of the near horizon of the black hole relative to the physics near the AdS$_3$ boundary.

\vfill

\noindent \today

\end{titlepage}
}

\newpage

\tableofcontents

\newpage
\section{Introduction}
AdS$_2$ quantum gravity plays an important role in our understanding of black holes. A prominent example is the construction of the quantum entropy function via AdS$_2$/CFT$_1$\cite{Sen:2008yk,Sen:2008vm},  which encodes classical and quantum properties of extremal black holes in agreement with our microscopic understanding in string theory \cite{Sen:2007qy,Mandal:2010cj,Sen:2014aja}. Unfortunately,  relative to higher dimensional instances of AdS/CFT,  we face some serious obstructions in building a holographic description of AdS$_2$.  One crucial obstacle is that  its symmetry prevents finite energy excitations, so capturing non-trivial dynamics requires a deformation that destroys the AdS$_2$ background \cite{Strominger:1998yg,Maldacena:1998uz}.

A proposal addressing this obstacle is  known as the nAdS$_2$/nCFT$_1$ correspondence \cite{Almheiri:2014cka,Maldacena:2016upp}. The first insights relied on studies of 2D models of gravity coupled to a scalar field (i.e., a dilaton), which are colloquially referred to as JT gravity \cite{Jackiw:1984je,Teitelboim:1983ux}. Some generalisations are those in  \cite{Grumiller:2015vaa,GrumillerMcNeesSalzerEtAl2017}. In these models the non-trivial profile of the dilaton breaks  explicitly the conformal symmetry of AdS$_2$, while being at the same time tied to the large diffeomorphisms at the boundary of AdS$_2$. Moreover, these diffeomorphisms induce an anomaly via a Schwarzian derivative. This symmetry breaking pattern is important: It governs the gravitational backreaction, such as the thermodynamic response and the quantum chaos characterising black holes. And so the persistent trend in nAdS$_2$ holography, coined with a `n' since we are `near' to our original configuration,  is that the deviations away from extremality are controlled by this pattern. For a  review see \cite{Sarosi:2017ykf}.

The application of this new framework to black hole physics has shown that, while JT models capture common features \cite{Jensen2016,Engelsoy:2016xyb,Almheiri:2016fws,Nayak:2018qej,Moitra:2018jqs,Moitra:2019bub}, the additional parameters for more general black holes display interactions that are not present in JT gravity \cite{Anninos:2017cnw,Kolekar:2018sba,Larsen:2018iou,Castro:2018ffi,Larsen:2018cts,Hong:2019tsx,Castro:2019crn}. This makes clear that there is new phenomena to be explored, that simpler models do not take into account. 

Our interest  therefore is to further explore the properties of nAdS$_2$/nCFT$_1$ with the goal of building a more refined understanding of the dynamics near the horizon of (near-)extremal black holes. We will revisit the renown BTZ black hole in AdS$_3$ gravity \cite{Banados:1992wn,Banados:1992gq} using the framework of nAdS$_2$ holography. We will treat the angular direction in BTZ as a compact direction along which we will dimensionally reduce to two dimensions. The resulting 2D theory of gravity contains, in addition to the metric, a gauge field and a dilaton as expected in Kaluza-Klein theory. 
Our work builds upon the developments in \cite{Almheiri:2016fws,Cvetic:2016eiv}, where the 2D holographic dictionary  was studied by dimensionally reducing the 3D Einstein-Hilbert action with a negative cosmological constant. Other relevant work includes \cite{Gaikwad:2018dfc}, which focuses on the effects of $U(1)$ Chern-Simons fields in 2D; see also \cite{Das:2017pif,Taylor:2017dly,Poojary:2018esz, Kolekar:2018sba, Kolekar:2018chf}. 

One question we investigate is the  relation between the nCFT$_1$, that describes the near horizon physics of near-extremal BTZ, to the parent CFT$_2$, that is dual to AdS$_3$. This relation is subtle: The conformal (Weyl) anomaly in AdS$_3$/CFT$_2$ can easily be confused with the anomaly appearing in nAdS$_2$/nCFT$_1$. Both are controlled by a Schwarzian derivative after all.\footnote{Related work that ties the Schwarzian derivative in the nCFT$_1$ to a Virasoro symmetry in a CFT$_2$ includes \cite{Turiaci:2017zwd,Mandal:2017thl,Mertens:2017mtv}.}  In the present paper we explore this relation by adding to the  Einstein-Hilbert action a gravitational Chern-Simons term: The resulting theory is topologically massive gravity (TMG). In the context of AdS$_3$/CFT$_2$, it is known that this theory contains both a conformal and gravitational anomaly, reflected in the boundary theory as a violation of parity that induces $c_L\neq c_R$ \cite{Kraus:2005zm}. Here $c_{L/R}$ are the left/right central charges in the CFT$_2$. 

Considering the gravitational Chern-Simons term adds a layer of complexity which, despite making some derivations more cumbersome, has several advantages. First, having a distinction between left and right movers will be particularly important when considering the thermodynamic responses in the nCFT$_1$, and its comparison to thermal properties of the CFT$_2$. Second,   
one of our main results is that the anomaly of nAdS$_2$/nCFT$_1$ is due to one chiral sector of the CFT$_2$, and hence it seems misleading to only discuss it as a Weyl anomaly.  The dilaton and gauge field will play a crucial role in this interpretation: Different choices of boundary conditions will impact the holographic interpretation we aim to build. Our strategy therefore will be to divide the analysis of holographic properties into two:
\begin{description}
\item[UV perspective:] This portion focuses on backgrounds in 2D that naturally uplift to asymptotically AdS$_3$ spacetimes. These are the running dilaton backgrounds in \cite{Cvetic:2016eiv}. Our emphasis here is to keep track  in the dimensionally reduced theory of the conformal and gravitational anomaly present in AdS$_3$/CFT$_2$. In this setup the reduced gravitational Chern-Simon term is somewhat dull: It modifies the conserved charges, but  disappears from the Ward identities in the lower dimensional theory.  
\item[IR perspective:] Here the starting point are solutions with a constant dilaton, leading to locally AdS$_2$ spacetimes. We coin these background IR, since they uplift to the near horizon physics of nearly extremal black holes.  We will then turn on a deformation  for the dilaton that ignites the key features of nAdS$_2$ holography. The asymptotic behaviour of the fields in this situation is different relative to the UV, and therefore changes various observables. In particular, the gravitational Chern-Simons term influences the anomalies appearing in nAdS$_2$/nCFT$_1$. 
\end{description}

Finally, the 2D theory we will consider contains higher derivative interactions, and hence encapsulates  a rich space of solutions.  Some related work that studies certain classes of solutions includes \cite{Grumiller:2003ad,Myung:2009sk,Alishahiha:2008rt}. Here we will exclusively focus on solutions of the 2D theory, that upon an uplift, can be interpreted as locally AdS$_3$ spacetimes; these are the solutions described in \cite{Cvetic:2016eiv}. This subsector is a consistent truncation of the theory, and it will suffice to explore dynamics related to the BTZ black hole. There are of course plenty of other interesting configurations, in particular warped AdS$_3$ black holes \cite{Moussa:2003fc,Bouchareb:2007yx,Anninos:2008fx}, which would be interesting to study in future work  using  the tools of nAdS$_2$ holography.

The paper is organised as follows: In Sec.\,\ref{sec:tmg} we will introduce the 3D parent theory, i.e. TMG, alongside with a review of the holographic properties, and summarise the thermodynamic effects of the gravitational Chern Simons term on the BTZ black hole. 
In Sec.\,\ref{sec:2DKK} we perform the dimensional reduction of TMG, and present its equations of motion in full generality. Sec.\,\ref{sec:UV} focuses on holographic renormalisation of the 2D theory with our UV perspective: After setting the appropriate boundary conditions in 2D, we evaluate the one-point functions and derive the renormalised action.  In Sec.\,\ref{sec:kk3d2d} we compare the 3D results in Sec.\,\ref{sec:hr3d} to our derivations in Sec.\,\ref{sec:HRuv}. This comparison illustrates that the 2D theory washes away some aspects of the gravitational anomaly, which we discuss. The results relevant to the near horizon physics of the BTZ are in Sec.\,\ref{sec:IR}. This is our IR setup, where the  starting point are AdS$_2$ backgrounds with a constant dilaton. We perform holographic renormalisation in nAdS$_2$, and already at early stages of the computations the differences with the UV become manifest, as we advertised above. Finally, in Sec.\,\ref{sec:schw} we make the symmetry breaking mechanism and anomalies in the 2D theory manifest. For this we derive the Schwarzian action for both the UV and IR perspective.  We discuss the interpolation between the UV and IR, and the role these anomalies have in the entropy of 2D black holes. App.\,\ref{app:btz} contains useful relations that cast the BTZ black hole as a 2D solution.

\section{Topologically massive gravity}\label{sec:tmg}

The addition of a gravitational Chern-Simons term to the Einstein-Hilbert action in three dimensions defines topologically massive gravity \cite{Deser:1981wh,Deser:1982vy,Deser:1991qk}.  The action is given by
\begin{align}\label{eq:TMG}
&I_{\scaleto{\text{3D}}{4pt}}= I_{\scaleto{\text{EH}}{4pt}}+ I_{\scaleto{\text{CS}}{4pt}}~,\cr
&I_{\scaleto{\text{EH}}{4pt}}=\frac{1}{16\pi G_3}\int \dd x^3\sqrt{-g}\left(R- 2\Lambda\right)~, \cr
&I_{\scaleto{\text{CS}}{4pt}}=  \frac{1}{32 \pi G_3 \mu}\int \dd x^3\sqrt{-g}\varepsilon^{MNL}\left(\Gamma^P_{M S}\partial_N \Gamma^S_{LP} + \frac{2}{3} \Gamma^P_{MS} \Gamma^S_{NQ}\Gamma^Q_{LP}\right)~, 
\end{align}
where we have included a negative cosmological constant $\Lambda= -1/\ell^2$, with $\ell$  the AdS$_3$ radius, and  $\mu$ is a real coupling with dimensions of mass; we are using convention where $\sqrt{-g}\,\varepsilon^{012}=-1$. There is also a gauge theory formulation of this theory, which uses a Chern-Simons description of 3D gravity plus a constraint \cite{Deser:1991qk, Carlip:2008qh,Chen:2011yx}.

The equations of motion of TMG read
\bea\label{eq:eomtmg}
R_{MN}-{1\over 2}g_{MN} R -{1\over \ell^2} g_{MN}=-{1\over \mu}C_{MN}~,
\eea
where $C_{MN}$ is the Cotton tensor,
\be\label{eq:cotton}
C_{MN}=\epsilon_M^{~~QP}\nabla_Q \left(R_{PN}-{1\over 4}g_{PN}R\right)~.
\ee
Note that the equations of motion are covariant, even though the action has explicit dependence on Christoffel symbols. It is also important to highlight that {\it all} locally AdS$_3$ spacetimes have vanishing Cotton tensor,  $C_{MN}=0$, which makes them automatically a solution to \eqref{eq:eomtmg}.

The novel solutions of TMG are those with $C_{MN}\neq 0$. An interesting subset of such solutions, denoted ``warped AdS$_3$,'' were constructed in \cite{Moussa:2003fc,Bouchareb:2007yx,Anninos:2008fx} along with warped  black hole counterparts; see also \cite{Nutku:1993eb,Gurses:2008wu,Chow:2009km,Anninos:2009jt}.  Viewed holographically, the main feature of asymptotically warped AdS$_3$ geometries is that they do not obey Brown-Henneaux boundary conditions  \cite{Brown:1986nw}.  Indeed, the nature and symmetries of their holographic descriptions is more intricate than those in AdS$_3$   \cite{Compere:2008cv,ElShowk:2011cm,Song:2011sr,Guica:2011ia,Detournay:2012pc}. Our focus here will be on locally AdS$_3$ configurations; we will postpone the study of warped AdS$_3$ spacetimes for future work.

\subsection{Holographic renormalisation}\label{sec:hr3d}

Some of the distinctive  properties of the gravitational anomaly  in TMG are very manifest in AdS$_3$/CFT$_2$ \cite{Deser:2002iw,Solodukhin:2005ns,Kraus:2005zm,Li:2008dq,Hotta:2008yq,Skenderis:2009nt}.  In this section we will provide a quick summary of the resulting boundary stress tensor for TMG, which is mainly based on  \cite{Kraus:2005zm,Skenderis:2009nt}. We will  focus only on locally AdS$_3$ solutions.\footnote{ At $\mu=1$, i.e. chiral gravity, there are some additional subtleties due to an additional logarithmic branch in the classical phase space. This introduces solutions that are not locally AdS, while still being asymptotical AdS for appropriate boundary conditions. We will not dwell with this special point, and instead refer the reader to \cite{Li:2008yz,Grumiller:2008qz,Maloney:2009ck}, and references within, for holographic properties of chiral gravity.}

 The application of Brown-Henneaux boundary conditions for TMG shows that the classical phase space of asymptotically AdS$_3$ (AAdS$_3$) backgrounds is organised in two copies of the Virasoro algebra with central charges
\be\label{eq:clcr}
c_R={3\ell\over 2G_3}\left(1-{1\over \mu\ell}\right) ~,\quad c_L= {3\ell\over 2G_3}\left(1+{1\over\mu\ell}\right)~.
\ee
This result uses the  Fefferman-Graham expansion of the 3D metric, which is given by 
\begin{align}\label{eq:FG3d}
 ds^2_3 &= \dd \eta^2 + g_{ij}(\eta,x)\dd x^i \dd x^j,\quad i,j\in\{0,1\} ~,\cr
g_{ij}(\eta,x)&=g_{ij}^{(0)}(x)\,e^{2\eta/\ell} + g_{ij}^{(2)}(x) + O (e^{-2\eta/\ell})~,
\end{align}
where as usual $x^i$ denotes  the boundary coordinates. In this context, the boundary stress tensor for the 3D theory is defined as the on-shell variation of the renormalised action with respect to the boundary metric 
\begin{align}\label{eq:boundarytensor3D}
\delta I_{\scaleto{\text{3D}}{4pt}}^{\scaleto{\text{ren}}{3pt}}=\frac{1}{2}\int\dd x^2\sqrt{g^{(0)}}\, T^{ij}\, \delta g_{ij}^{(0)}~.
\end{align}
Here $I_{\scaleto{\text{3D}}{4pt}}^{\scaleto{\text{ren}}{3pt}}$ contains, in addition to \eqref{eq:TMG}, the Gibbons-Hawking-York term, and a boundary cosmological constant, i.e. the standard counterterms in the holographic renormalisation of the 3D Einstein-Hilbert action with AAdS$_3$  boundary conditions \cite{Balasubramanian:1999re,deHaro:2000vlm,Skenderis:2002wp}. One very interesting aspect of TMG is that the gravitational Chern-Simons term does not lead to new divergences: the variation with respect to $g_{ij}^{(0)}$ of $I_{\scaleto{\text{CS}}{4pt}}$ is finite as $\eta\to \infty$. There are however some ambiguities in the variation of $I_{\scaleto{\text{CS}}{4pt}}$, due to the choice of renormalisation scheme in theories with a gravitational anomaly; we will review those choices in the following.

\paragraph{Consistent stress tensor \cite{Skenderis:2009nt}.} The stress tensor that arises from a well defined variational principle is given by
\begin{equation} 
T_{ij}= \frac{1}{8 \pi G_3 \ell }\left(g_{ij}^{(2)}-g_{ij}^{(0)}g_{kl}^{(2)}g_{(0)}^{kl}\right) - \frac{1}{16\pi G_3 \mu \ell^2}\left(g_{ik}^{(2)}\varepsilon_{lj}g_{(0)}^{kl} +g_{jk}^{(2)}\varepsilon_{li}g_{(0)}^{kl} \right) -{1\over 16\pi G_3\mu} A_{ij}~, \label{eq:STvR}
\end{equation}
where the additional term 
 \begin{align}
 A_{ij}= \frac{1}{4}\varepsilon^{kl}D_k \partial_l g_{ij}^{(0)}-\frac{1}{8}\varepsilon_i^{~ k}\varepsilon_j^{~l}\varepsilon^{mn}D_l\partial_m g_{nk}^{(0)}-\frac{1}{8}\varepsilon_j^{~ k}\varepsilon_i^{~l}\varepsilon^{mn}D_l\partial_m g_{nk}^{(0)}~,
 \end{align}
solely depends on the boundary metric $g_{ij}^{(0)}$. 
Here $\varepsilon_{ij}$ is the epsilon tensor for the boundary metric, and we set $\sqrt{-g^{(0)}}\varepsilon^{01}=-1$; $D_i$ is the covariant derivative with respect to $g^{(0)}_{ij}$.

The trace anomaly and Ward identity for this form of the stress tensor read
\begin{align}
T_{~i}^{i}=\frac{c}{24\pi} R^{(0)} +\frac{\bar{c}}{12\pi}A_{~i}^i~,\cr 
 D_i T^{ij}= -\frac{\bar{c}}{24\pi}g_{(0)}^{ij}\varepsilon^{kl}\partial_m\partial_k\Gamma_{il}^m~,\label{eq:tracean}
\end{align}
where $R^{(0)}$ denotes the Ricci scalar for $g_{ij}^{(0)}$ and we also used \eqref{eq:clcr} and introduced $c=(c_L+c_R)/2$ and $\bar c=(c_L-c_R)/2$. Casting the diffeomorphism anomaly as in \eqref{eq:tracean} is in accordance with the Wess-Zumino consistency conditions, albeit the expressions are not covariant. The lack of covariance is reflected on the failure  of $A_{ij}$ to be a tensor.

\paragraph{Covariant  stress tensor \cite{Kraus:2005zm,Solodukhin:2005ah}.} The term $A_{ij}$ in the stress tensor \eqref{eq:STvR} does not carry information that depends on the ``state'', i.e. it does not depend on $g^{(2)}_{ij}$. If one removes $A_{ij}$, the resulting holographic stress tensor reads
\begin{align} 
t_{ij}= \frac{1}{8 \pi G_3 \ell }\left(g_{ij}^{(2)}-g_{ij}^{(0)}g_{kl}^{(2)}g_{(0)}^{kl}\right) - \frac{1}{16\pi G_3 \mu \ell^2}\left(g_{ik}^{(2)}\varepsilon_{lj}g_{(0)}^{kl} +g_{jk}^{(2)}\varepsilon_{li}g_{(0)}^{kl}\right)~.\label{eq:STTMG1}
\end{align}
The trace anomaly and Ward identity now are
\begin{align}\label{eq:trace2}
t_{~i}^i=\frac{c}{24\pi} R^{(0)}~, \qquad   D_j t^{ij}=\frac{\bar{c}}{24\pi}\varepsilon^{ij}\partial_j R^{(0)}~.
\end{align}
In contrast to \eqref{eq:tracean}, these expressions are covariant with respect to the boundary metric, which makes this stress tensor receive the name `covariant'. The sacrifice here is that it does not satisfies the Wess-Zumino conditions. 

\paragraph{Conserved and Lorentz violating stress tensor \cite{Solodukhin:2005ah}.}
Finally, one can also insist that the stress tensor is conserved. From \eqref{eq:trace2} we see that this is easily achieved by defining
\begin{align}\label{eq:t3}
\hat{t}_{ij}= t_{ij}+\frac{\ell}{16 \pi G_3\mu}\varepsilon_{ij}R^{(0)}~.
\end{align}
However, now we have an object that is not symmetric, which is a significant sacrifice in this definition. From here it is natural to cast $\hat{t}^{\hat{a}}_{~i}= e^{~\hat{a}}_j t_{~i}^{j}$, where $\hat{t}^{\hat{a}}_{~i}$ is the response of the action to variations of the vielbeins $e^{~\hat{a}}_i$. Note that $\hat{t}^{\hat{a}}_{~i}$ is also not invariant under local Lorentz transformations. At the price of loosing Lorentz invariance, the relevant identities for \eqref{eq:t3} are
\begin{align}\label{eq:456}
\varepsilon_{\hat{a}}^{~i}\,\hat{t}^{\hat{a}}_{~i}= \frac{c}{12 \pi}R^{(0)}~, \quad \quad
D^j \hat{t}^{\hat{a}}_{~j}=0~.
\end{align}
We conclude this section by noting, that apart from theories with a gravitational anomaly, one encounters asymmetric boundary stress tensors in non-relativistic theories.  In a holographic context, see for example \cite{Ross:2011gu}.

\subsection{BTZ black hole}\label{sec:BTZ}

In this section we introduce the BTZ black hole in TMG and review some of its thermodynamic properties. 
The metric of the rotating BTZ solution is \cite{Banados:1992gq} 
\begin{align}
ds^2_3 = -\frac{(\rho^2-\rho_+^2)(\rho^2-\rho_-^2)}{\ell^2\rho^2}\dd t^2 + \frac{\ell^2 \rho^2}{(\rho^2-\rho_+^2)(\rho^2 - \rho_-^2)}\dd \rho^2 + \rho^2\left(\dd \varphi - \frac{\rho_+ \rho_-}{\ell \rho^2}\dd t\right)^2~, \label{eq:BTZ}
\end{align}
where $\rho_\pm$ are the position of the outer/inner horizon; without loss of generality, we will pick $\rho_+>\rho_->0$. In the absence of the gravitational Chern-Simons term, mass and angular momentum are given by
\begin{align}\label{eq:chargesEH}
{m}= \frac{\rho_+^2 + \rho_-^2}{8 G_3 \ell^2}~, \qquad j= \frac{\rho_+ \rho_-}{4 G_3\ell}~.
\end{align}
 The additional Chern Simons term contributes to the conserved charges in TMG. In particular the gravitational mass and angular momentum  read
\begin{align} \label{eq:chargeTMG}
M&= \frac{1}{24 \ell^3}\left( c_R\,(\rho_+ + \rho_-)^2 + c_L\,(\rho_+ - \rho_-)^2\right) ~,\cr
J&=  \frac{1}{24 \ell^2}\left(c_R\,(\rho_+ + \rho_-)^2- c_L\,(\rho_+- \rho_-)^2\right)~.
\end{align}
It is worth noting that  all variants of the boundary stress tensor presented above --i.e. \eqref{eq:STvR}, \eqref{eq:STTMG1} and \eqref{eq:t3}-- report the same answer for $M$ and $J$.  It is also instructive to relate the charges in TMG to those in \eqref{eq:chargesEH} 
\begin{align}
M \ell - J&=\left(1 + \frac{1}{\mu \ell}\right)\le(m \ell- j\ri)~, \cr
M \ell + J&=\left(1 - \frac{1}{\mu \ell}\right)\le(m \ell + j\ri)~.
\end{align}


\subsubsection{Thermodynamics near Extremality}\label{sec:btzthermo}

An important component of our holographic analysis of nAdS$_2$  encompasses the thermodynamic response in the presence of an irrelevant deformation. In the context of the 3D BTZ black hole this would correspond to the entropy near-extremality, which we review here.  More details on this limit are presented in App.\,\ref{app:btz}.

The Wald entropy of the BTZ black hole in TMG receives a non-trivial contribution which has been well documented and studied in \cite{Saida:1999ec, Kraus:2005vz, Solodukhin:2005ns,Sahoo:2006vz,Tachikawa:2006sz, Gupta:2008ki,Gupta:2008ki,Detournay:2012ug}. The resulting expression is
\begin{align}\label{eq:wald}
S&=\frac{\pi}{6\ell}\left(c_L\,(\rho_+ - \rho_-)+ c_R\,(\rho_+ + \rho_-)\right)~. 
\end{align}
Using the expression for mass and angular momentum in \eqref{eq:chargeTMG}, it is straight forward to verify the first law
\be
dM = T dS - \Omega\, d J~,
\ee
where the temperature and angular velocity are
\begin{align}\label{eq:mjT}
T=\frac{\rho_+^2 - \rho_-^2}{2\pi \ell^2 \rho_+}~,\qquad  \Omega=- {\rho_-\over \ell \rho_+}~.
\end{align}
Note that these potentials are independent of the gravitational couplings, $G_3$ and $\mu$, as expected since they are completely determined by the Euclidean regularity of the line element \eqref{eq:BTZ}. 

At extremality we have $\rho_+=\rho_-\equiv \rho_0$. In this limit it follows from \eqref{eq:mjT} that the temperature is zero, while the mass and entropy are
\be\label{eq:112}
M_{\rm ext}= \frac{c_R}{6 \ell^3} \,\rho_0^2~,\qquad S_{\rm ext}= 2\pi \sqrt{ {c_R\over 6} M_{\rm ext}\ell} ~.
\ee
 Near extremality is a small deviation of $\rho_+$ away from its extremal value $\rho_0$, i.e. $\rho_+= \rho_0+ \delta$ with $\delta$ a small parameter. In particular, we will deviate from extremality such that we increase the temperature $T$ slightly away from zero, which increases the mass $M$ of the black hole while keeping the angular momentum $J$  fixed.  The implementation of this limit gives a mass increase by
\begin{align}
\Delta E =M- M_{\rm ext}= \frac{1}{M_{\mathrm{gap}} }T^2 + \ldots~,  \label{eq:diffM}
\end{align}
where the dots indicate that this is an expansion around small values of $T$. The response of the mass in this limit  is quadratic with $T$ as expected \cite{Almheiri:2016fws}, where the coefficient that relates them is the mass gap
\begin{align}\label{113}
M_{\mathrm{gap}}= \frac{8 G_3}{\pi^2 \ell^2}\frac{1}{\left(1 + \frac{1}{\mu \ell}\right)} = \frac{12 }{\pi^2\ell c_L}~.
\end{align}
It follows that the response of the entropy \eqref{eq:wald} near extremality is linear in the temperature
\begin{align}\label{114}
S= S_{\mathrm{ext}} + \frac{2}{M_{\mathrm{gap}}} T + \ldots~. 
\end{align}

It is useful to cast these expressions in the language of the dual CFT$_2$. In this context the entropy \eqref{eq:wald} can be identified with the density of states distinctive of the Cardy regime,
\be
S = S_L+S_R~,
\ee
where the contribution to the entropy splits into a left and right moving part given by
\be
S_{L/R}= 2\pi \sqrt{{c_{L/R}\over 6} \,h_{L/R}}~, \qquad h_{L/R}= {1\over 2}(M\ell \mp J)~.
\ee
At extremality, we have $S_L=0$ and $S_R= S_{\rm ext}$. The first deviation away from extremality in \eqref{114} is due to the response of $S_L$, while $S_R$ remains dormant.  The addition of the gravitational Chern-Simons term gives a way to disentangle the role of right versus left degrees of freedom in the CFT$_2$.  And the interpretation is rather clear: The right movers control the ground state degeneracy at zero temperature, while the excitations near extremality are governed by the left moving excitations.


\section{2D Theory}\label{sec:2DKK}

In this section we describe the  2D theory obtained via a dimensional reduction of \eqref{eq:TMG}. The ansatz for the 3D metric is
\begin{align}
ds^2_{3} = g_{MN} \dd x^M \dd x^N =g_{\mu\nu}\dd x^\mu \dd x^\nu + e^{-2\phi}\left(\dd z+ A_\mu \dd x^\mu\right)^2~. \label{eq:KKansatz}
\end{align}
Here $z$ is a compact direction with period $2\pi L$; the Greek indices run along the two dimensional directions, $\mu,\nu=0,1$. From the two dimensional perspective, $g_{\mu\nu}$ is the metric, $A_\mu$ is a gauge field and $\phi$ will be interpreted as the dilaton field. 

The Kaluza-Klein reduction of $I_{\scaleto{\text{3D}}{4pt}}$, while tedious, is straight forward. The resulting action is \cite{Guralnik:2003we,Sahoo:2006vz} 
\begin{align}\label{eq:2Daction}
I_{\scaleto{\text{2D}}{4pt}} = I_{\scaleto{\text{EMD}}{4pt}} + I_{\scaleto{\text{rCS}}{4pt}}~,
\end{align}
where the first term, coming from the Einstein-Hilbert piece in \eqref{eq:TMG}, reads
\begin{align}\label{eq:einsteinmaxwell}
I_{\scaleto{\text{EMD}}{4pt}}= \frac{L}{8 G_3}\int \dd^2 x\,\sqrt{-g}\,e^{-\phi}\left(R + \frac{2}{\ell^2}- \frac{1}{4}e^{-2\phi}\, F_{\mu\nu}F^{\mu\nu}\right) ~,
\end{align}
and the piece related to the gravitational Chern-Simons theory is
\begin{align}\label{eq:KKreducedCS}
I_{\scaleto{\text{rCS}}{4pt}}&= \frac{L}{32G_3\mu}\int\dd^2x\,e^{-2\phi}\epsilon^{\mu\nu}\left(F_{\mu\nu}R + F_{\mu\rho}F^{\rho\sigma}F_{\sigma\nu}\,e^{-2\phi} - 2F_{\mu\nu} D^2\phi\right)~. 
\end{align}
Here $\epsilon^{\mu\nu}$ is the epsilon symbol, where $\epsilon^{01}=1$, and  $D_\mu$ is the covariant derivative with respect to the two dimensional metric  $g_{\mu\nu}$. 

In the following, we will refer to \eqref{eq:einsteinmaxwell} as the Einstein-Maxwell-Dilation theory (EMD), which captures the two derivative dynamics of the dimensional reduction. The action \eqref{eq:KKreducedCS} will be denoted as a reduced-Chern-Simons term (rCS), which contains the dynamics due to the 3D gravitational anomaly.  As observed in \cite{Sahoo:2006vz}, it is interesting to note that \eqref{eq:KKreducedCS} is gauge and diffeomorphism invariant; this is related to the fact that the 3D equations of motion \eqref{eq:eomtmg} are diffeomorphism invariant. 

The  equations of motion  are
\begin{align}\label{eq:MaxDil}
&\epsilon^{\alpha\beta}\partial_\beta \left(e^{-3\phi}f +\frac{1}{2\mu}\,e^{-2\phi}\left(R  + 3\,e^{-2\phi}f^2 - 2 D^2 \phi \right)\right)=0~, \cr
 &e^{-\phi}\left(R + \frac{2}{\ell^2}+ \frac{3}{2}\,e^{-2\phi}f^2\right) + \frac{1}{\mu}\,e^{-2\phi}f\left(R +2\,e^{-2\phi}f^2 - 2 D^2 \phi\right)+ \frac{1}{\mu} D^2\left(e^{-2\phi}f\right)=0~,
 \end{align}
 which are the Maxwell and dilaton equations respectively. The variation with respect to the metric gives
 \begin{align}\label{eq:einstein}
&g_{\alpha\beta}\left(D^2e^{-\phi} - \frac{1}{\ell^2}\,e^{-\phi} + \frac{1}{4}\,e^{-3\phi}f^2\right)-D_\alpha D_\beta e^{-\phi} 
\cr&\quad+\frac{1}{2\mu}\Big((D_{\alpha}e^{-2\phi}f) D_{\beta}\phi +(D_{\beta}e^{-2\phi}f) \DD_{\alpha}\phi 
- D_\alpha D_\beta (e^{-2\phi}f) \Big)\cr
&\quad+\frac{1}{2\mu}g_{\alpha\beta} \Big({1\over 2}\, e^{-2\phi} f R-e^{-2\phi}f D^2\phi -D_\mu(e^{-2\phi}f) D^\mu\phi +D^2(e^{-2\phi}f) + e^{-4\phi}f^3  \Big)=0~.
\end{align}
It is also useful to record the trace of Einstein's equation, which reads
\begin{align}\label{eq:eomtrace}
 D^2 e^{-\phi}- \frac{2}{\ell^2}\,e^{-\phi} + \frac{1}{2}\,e^{-3\phi}f^2 + \frac{1}{2\mu}\,e^{-2\phi}f\left(R +2\,e^{-2\phi}f^2 - 2 D^2 \phi\right)+ \frac{1}{2\mu} D^2\left(e^{-2\phi}f\right)=0~.
\end{align}
In the above equations we introduced\footnote{In terms of the epsilon tensor, $\varepsilon_{\alpha \beta}=\sqrt{-g}\,\epsilon_{\alpha\beta}$, we have
\be
f=-\frac{1}{2}\varepsilon^{\alpha\beta}F_{\alpha\beta}~,\qquad F_{\alpha\beta}=\varepsilon_{\alpha\beta} f~,\nonumber
\ee
where $\varepsilon^{\alpha\beta}\varepsilon_{\alpha\beta}=-2$.
} 
\begin{align}
f \equiv \frac{1}{2\sqrt{-g}}\,\epsilon^{\alpha\beta}F_{\alpha\beta}~,
\end{align}
which transforms as a scalar under diffeomorphisms. It is important to emphasise that \eqref{eq:2Daction} is a consistent truncation of TMG: All solutions to the equations of motion \eqref{eq:MaxDil}-\eqref{eq:einstein}, when uplifted via \eqref{eq:KKansatz}, are solutions to \eqref{eq:eomtmg}.


\section{Holographic renormalisation: UV perspective}\label{sec:UV}

One of our goals is to capture holographic properties of the 2D theory \eqref{eq:2Daction}. We will start by considering backgrounds that have a running dilaton profile. In particular we will impose boundary conditions on the 2D fields that, upon an uplift to 3D, are interpreted as asymptotically AdS$_3$ backgrounds. For this reason, we coin this section a UV perspective to holographic renormalisation. 

\subsection{Background solution}\label{sec:solUV}

To characterise the space of solutions, we will use throughout the gauge 
\be\label{eq:gaugechoice}
ds^2= \dd r^2+ \gamma_{tt}\, \dd t^2 ~, \qquad A_r =0~.
\ee
Our interest here will be restricted to a very specific class of solutions: backgrounds that solve the equations of motion of $I_{\scaleto{\text{EMD}}{4pt}}$. As in the three dimensional parent theory, any solution to $I_{\scaleto{\text{EMD}}{4pt}}$ will be a solution to $I_{\scaleto{\text{rCS}}{4pt}}$. The most general solutions to EMD where constructed in \cite{Cvetic:2016eiv}, which we briefly summarise here.  In EMD the  gauge field is fixed to
\be\label{eq:FEMD}
F_{rt}=-2Q \, e^{3\phi} \sqrt{-\gamma}~.
\ee
Solutions with a non-constant dilaton profile satisfy
\be\label{eq:runningG}
\sqrt{-\gamma}=\frac{\alpha(t)}{\lambda'(t)}\,\partial_te^{-\phi}~, 
\ee
where the dilaton is 
\begin{align} \label{eq:runningD}
e^{-2\phi}= \lambda(t)^2\,e^{2r/\ell}\left(1 + {\ell^2\over2\lambda(t)^2}\,\mathfrak{m}(t)\,e^{-2r/\ell} + \frac{\ell^2}{16\lambda(t)^4}\left(\ell^2\,\mathfrak{m}(t)^2- 4Q^2\right)e^{-4r/\ell}\right)~,
\end{align}
and we introduced
\be\label{eq:defm123}
\mathfrak{m}(t)\equiv m_0-\left(\frac{\lambda'(t)}{\alpha(t)}\right)^2~.
\ee
Here $\alpha(t)$ and $\lambda(t)$ are arbitrary functions of time that will be identified with the sources for the metric and dilaton, respectively; $m_0$ and $Q$ are constants.   For the subsequent analysis it will be useful to record the asymptotic behaviour of the solutions, which reads
\begin{align} \label{eq:asymptoticvalues}
e^{-2\phi}&= \lambda^2e^{2r/\ell}+ \frac{\ell^2}{2}\mathfrak{m}+ O(e^{-2r/\ell})~,\cr
\gamma_{tt}&= -\alpha^2e^{2r/\ell}+\frac{\ell^2\alpha^2}{2\lambda}\left(\frac{\mathfrak{m}}{\lambda}-{\mathfrak{m}'\over \lambda'}\right)+ O(e^{-2r/\ell})~,\cr
A_t& = \nu+\frac{\ell \alpha}{\lambda^3}\, Q \, e^{-2r/\ell}  +O(e^{-4r/\ell})~.
\end{align}
The radial dependence of the gauge field $A_t$ is fixed by \eqref{eq:FEMD}; its source is $\nu(t)$, which is locally pure gauge.  To be concise we have omitted the explicit time dependence of $\alpha(t), \lambda(t)$, $\mathfrak{m}(t)$ and $\nu(t)$, and denoted time derivatives with a prime. 
The important feature here, to be contrasted with the asymptotic behaviour in Sec.\,\ref{sec:IR}, is that the gauge field has a sub-leading behaviour relative to the dilaton and 2D metric.
 
\subsection{Renormalised observables}\label{sec:HRuv}
An important portion of performing holographic renormalisation is to obtain finite variations of the action provided a set of boundary conditions.\footnote{ We are presenting here a very concise view on holographic renormalisation. For a current overview of the subject we refer to \cite{Papadimitriou:2010as,ElvangHadjiantonis2016}.} In this section we will impose boundary conditions compatible with  the leading behaviour in \eqref{eq:asymptoticvalues},  and require that the renormalised on-shell variation of the action 
\begin{align}\label{eq:varEMDUV1}
\delta I_{\scaleto{\text{2D}}{4pt}}^{\scaleto{\text{UV}}{4pt}}= \int \dd t \left( \Pi^{tt}\,\delta\gamma_{tt} + \Pi_\phi \,\delta \phi + \Pi^t  \,\delta A_t\right)~,
\end{align}
remains finite and integrable. More explicitly, starting from the bulk action \eqref{eq:2Daction}, we will add boundary terms leading to a functional $I_{\scaleto{\text{2D}}{4pt}}^{\scaleto{\text{UV}}{4pt}}$ whose variations are finite when 
\be\label{eq:bc1}
\delta\gamma_{tt}= -2\,\alpha\, e^{2r/\ell}\, \delta \alpha~, \quad \delta e^{-\phi} = e^{r/\ell} \delta \lambda~, \quad \delta A_t =\delta \nu~,
\ee
as we take $r\to \infty$.
These are our UV boundary conditions.  In terms of these sources, we have
\begin{align}\label{eq:varEMDUV}
\delta I_{\scaleto{\text{2D}}{4pt}}^{\scaleto{\text{UV}}{4pt}}= \int \dd t \le(\mathcal{T}_{\scaleto{\text{UV}}{4pt}} \,\delta \alpha + {\alpha\over \lambda} \,\mathcal{O}_{\scaleto{\text{UV}}{4pt}}\, \delta \lambda -  \alpha\mathcal{J}^t_{\scaleto{\text{UV}}{4pt}} \,\delta\nu \ri)~,
\end{align}
where we have introduced the one-point functions conjugate to each source. The relation to the momenta variables in \eqref{eq:varEMDUV1}  is given by 
\begin{align}
&\mathcal{T}_{\scaleto{\text{UV}}{4pt}}\equiv\frac{2}{\alpha}\lim_{r\rightarrow\infty}\Pi_{t}^{~t}~,\quad
\mathcal{O}_{\scaleto{\text{UV}}{4pt}}\equiv -\frac{1}{\alpha} \lim_{r\rightarrow\infty}\Pi_{\phi}\label{eq:OrCSHR}~,\quad
\mathcal{J}^t_{\scaleto{\text{UV}}{4pt}}\equiv -\frac{1}{\alpha}\lim_{r\rightarrow\infty}\Pi^t~.
\end{align}

Recall that our action has a contribution from the EMD action \eqref{eq:einsteinmaxwell} and the rCS action in \eqref{eq:KKreducedCS}. 
Holographic renormalisation for EMD, with the boundary conditions \eqref{eq:bc1}, was done in detail in \cite{Cvetic:2016eiv}, and  we will just highlight the main results. Varying the action  \eqref{eq:einsteinmaxwell} by itself  leads to well known pathologies. These are cured by addition of the Gibbons-Hawking term 
\begin{align}
I_{\scaleto{\text{GH}}{4pt}}&= \frac{L}{4G_3}\int\dd t\,\sqrt{-\gamma}\, e^{-\phi} K \label{eq:GHEMD}~,
\end{align}
which leads to Dirichlet boundary conditions on the metric, and the counterterm\footnote{As observed in  \cite{Cvetic:2016eiv}, in \eqref{eq:IcEMD} there is an additional term due to the conformal anomaly. It is a total derivative so it won't contribute to \eqref{eq:OrCSHR} and it will be ignored in the following.} 
\begin{align}\label{eq:IcEMD}
I_{\scaleto{\text{c1}}{4pt}}&=-\frac{L}{4G_3\ell}\int \dd t\, \sqrt{-\gamma}\,e^{-\phi}~,
\end{align}
that renders the variation of the action finite for \eqref{eq:bc1}.
In \eqref{eq:GHEMD}  $K$ is the trace of the  extrinsic curvature, which for our choice of gauge in \eqref{eq:gaugechoice} reads $K=\partial_r\log\sqrt{-\gamma}$.
The renormalised action is then 
\begin{align}
I_{\scaleto{\text{EMD}}{4pt}}^{\scaleto{\text{ren}}{3pt}}= I_{\scaleto{\text{EMD}}{4pt}}+ I_{\scaleto{\text{GH}}{4pt}}+ I_{\scaleto{\text{c1}}{4pt}}~.
\end{align}
 The variation of $I_{\scaleto{\text{EMD}}{4pt}}^{\scaleto{\text{ren}}{3pt}}$ results in the renormalised canonical momenta
\begin{align} \label{eq:PittEMD}
\delta I_{\scaleto{\text{EMD}}{4pt}}^{\scaleto{\text{ren}}{3pt}}=  &\frac{L}{8G_3} \int \dd t \sqrt{-\gamma}\left( \partial_re^{-\phi}- \frac{1}{\ell}e^{-\phi}\right) \gamma^{tt} \delta\gamma_{tt} \cr
&+ \frac{L}{4G_3}\int \dd t \sqrt{-\gamma} \left(K- \frac{1}{\ell}\right)e^{-\phi} \delta \phi -\frac{L}{4G_3} \int \dd t \, Q \, \delta A_t ~.
\end{align}

The contribution of the rCS action to \eqref{eq:varEMDUV1} is rather interesting. The on- shell variation of $I_{\scaleto{\text{rCS}}{4pt}}$ leads to
\begin{align}
\delta I_{\scaleto{\text{rCS}}{4pt}}=&-\frac{L}{16G_3 \mu}\int \dd t\,e^{-2\phi}\left(R + 12 Q^2\, e^{4\phi}-2D^2\phi \right)\delta A_t~\cr 
&-\frac{L}{4G_3 \mu}\int \dd t  \, Q \,e^{\phi} \delta \le(\sqrt{-\gamma} K\ri)  +\frac{L}{4G_3 \mu}\int \dd t \sqrt{-\gamma}\, Q\le((\partial_re^{\phi})\delta \phi - e^{\phi}\, \delta\le( \partial_r \phi\ri) \right)~, \label{eq:varIrCS}
\end{align}
where we used \eqref{eq:FEMD} to simplify this expression. This variation does not lead to divergences for \eqref{eq:bc1} and falls off in \eqref{eq:asymptoticvalues}, in accordance with the variation of the graviational Chern-Simons term in AAdS$_3$ spacetimes.  Also, not surprisingly, we find variations of derivatives of the metric and dilaton. In order to restore Dirichlet boundary conditions for these fields, we add  two extrinsic boundary terms:
\begin{align}\label{eq:ctermuv}
I_{\scaleto{\text{c2}}{4pt}}=\frac{L}{4G_3\mu }\int \dd t\sqrt{-\gamma}\, Q\, e^{\phi}\, K + \frac{L}{4G_3\mu}\int \dd t \sqrt{-\gamma}\, Q \, \partial_ r e^{\phi}~.
\end{align}
With this we obtain
\begin{align}
\delta \le(I_{\scaleto{\text{rCS}}{4pt}} + I_{\scaleto{\text{c2}}{4pt}}\ri)=&-\frac{L}{16G_3 \mu}\int \dd t~ e^{-2\phi}\left(R + 12 Q^2 e^{4\phi}-2D^2\phi \right)\delta A_t~\cr 
&+\frac{L}{8G_3 \mu}\int \dd t \sqrt{-\gamma} \, Q \, (\partial_ r e^{\phi})\, \gamma^{tt} \delta\gamma_{tt} \cr&+\frac{L}{4G_3 \mu}\int \dd t \sqrt{-\gamma} \, Q\, \le(2\,\partial_r e^{\phi} + K e^\phi\ri) \delta \phi \,. \label{eq:varfinal}
\end{align}
Gathering our contributions from \eqref{eq:PittEMD} and \eqref{eq:varfinal} and using \eqref{eq:asymptoticvalues} the renormalised one-point functions are
\begin{align}\nonumber
\mathcal{T}_{\scaleto{\text{UV}}{4pt}}&=  -\frac{L \ell}{8G_3} \frac{ \mathfrak{m}}{\lambda}  -\frac{L}{4 G_3\mu\ell}\frac{Q }{\lambda}~,\\ \nonumber
\mathcal{O}_{\scaleto{\text{UV}}{4pt}}&=\frac{L \ell }{8G_3}\left(\frac{\mathfrak{m}}{\lambda}- \frac{\mathfrak{m}'}{\lambda'} \right) +\frac{L}{4 G_3\mu\ell}\frac{Q }{\lambda}~,\\ \label{eq:cch1}
\mathcal{J}^t_{\scaleto{\text{UV}}{4pt}}&= \frac{L}{4G_3}\frac{Q}{\alpha}+ \frac{L}{8 G_3\mu}\frac{m_0}{\alpha}~
\end{align}
and the renormalised on-shell boundary action is 
\begin{align}\label{eq:uvrenac}
I_{\scaleto{\text{2D}}{4pt}}^{\scaleto{\text{UV}}{4pt}}&=I^{\scaleto{\text{ren}}{3pt}}_{\scaleto{\text{EMD}}{4pt}}+I_{\scaleto{\text{rCS}}{4pt}}+I_{\scaleto{\text{c2}}{4pt}}\cr
&= -\frac{L}{8G_3}\ell\int \dd t ~\left(\frac{\alpha}{\lambda} \le[m_0 +2{Q\over \mu \ell^2}\ri]+ \frac{\lambda'^2}{\alpha\lambda}+ \frac{2\nu}{\ell} \le[Q + {m_0\over 2 \mu}\ri]\right)~.
\end{align}
The above boundary action clearly satisfies
\begin{align}\label{eq:uvOp}
\mathcal{T}_{\scaleto{\text{UV}}{4pt}}&= \frac{\delta I_{\scaleto{\text{2D}}{4pt}}}{\delta\alpha}~, \quad \mathcal{O}_{\scaleto{\text{UV}}{4pt}} = \frac{\lambda}{\alpha}\frac{\delta I_{\scaleto{\text{2D}}{4pt}}}{\delta \lambda}~, \quad \mathcal{J}^t_{\scaleto{\text{UV}}{4pt}}= -\frac{1}{\alpha}\frac{\delta I_{\scaleto{\text{2D}}{4pt}}}{\delta\nu}~.
\end{align}
At this stage, the effect of adding rCS to EMD is to shift $m_0 \to m_0 +2{Q\over \mu \ell^2}$ and $Q\to Q + {m_0\over 2 \mu}$.
Additionally we observe that (\ref{eq:cch1}) obeys
\begin{align}
\mathcal{T}_{\scaleto{\text{UV}}{4pt}}+\mathcal{O}_{\scaleto{\text{UV}}{4pt}}= \frac{L\ell}{4G_3}\frac{1}{\alpha}\partial_t\left(\frac{\lambda'}{\alpha}\right)~,
\end{align}
and
 \begin{align}\label{eq:conservedcurrents}
&\partial_t\mathcal{J}^t_{\scaleto{\text{UV}}{4pt}}+\mathcal{J}^t_{\scaleto{\text{UV}}{4pt}}\,\partial_t\log\alpha= 0~,\quad 
\partial_t\mathcal{T}_{\scaleto{\text{UV}}{4pt}}-\mathcal{O}_{\scaleto{\text{UV}}{4pt}}\,\partial_t\log\lambda=0~.
 \end{align}

\subsection{KK reduction of AdS$_3$/CFT$_2$}\label{sec:kk3d2d}

In this last portion we will dimensionally reduce the different boundary stress tensors in 3D of  Sec.\,\ref{sec:hr3d}, and contrast them against the 2D quantitites in  Sec.\,\ref{sec:HRuv}. For the two derivative theory in 3D and 2D this was done in \cite{Cvetic:2016eiv};  this is also similar to the analysis done for Lifshitz holography in \cite{Christensen:2013rfa}, which relates currents and Ward identities via dimensional reduction from 5D down to 4D.

To start, we will first relate the 3D quantities to our  2D variables.   From \eqref{eq:KKansatz} we have 
\begin{align}\label{eq:map23}
g_{tt}= \gamma_{tt}+ e^{-2\phi}A_t^2~, \quad g_{tz}= e^{-2\phi}A_t~,\quad g_{zz}=e^{-2\phi}~,
\end{align}
 which in turn  implies that the boundary metric of the Fefferman-Graham expansion in \eqref{eq:FG3d} reads
\begin{align} \label{eq:g0}
g_{ij}^{(0)}= \begin{pmatrix}
\lambda^2\nu^2-\alpha^2 & \lambda^2 \nu \\
\lambda^2 \nu & \lambda^2
\end{pmatrix}~,
\end{align}
where we used \eqref{eq:asymptoticvalues}. In a similar fashion  we can read off $g_{ij}^{(2)}$ :
\begin{align}\label{eq:g2}
g^{(2)}_{tt}&=\frac{\ell^2}{2}\,\mathfrak{m}\left(\nu^2+\frac{\alpha^2}{\lambda^2}\right)+2\,\ell\,Q\,\frac{\alpha\,\nu}{\lambda}- \frac{\ell^2}{2}\,\frac{\alpha^2}{\lambda\,\lambda'}\,\mathfrak{m}'~,\cr
g^{(2)}_{zz}&= \frac{\ell^2}{2}\,\mathfrak{m}~, \quad  
g^{(2)}_{tz}=\ell\,Q\,\frac{\alpha}{\lambda}+ \frac{\ell^2}{2}\,\nu\,\mathfrak{m}~.
\end{align} 
With these we will relate variations of the action with respect to $g_{ij}^{(0)}$ to variations with respect to $\alpha$, $\lambda$ and $\nu$. This leads to
\be\label{eq:var23}
\frac{1}{2}\int\dd x^2\sqrt{g^{(0)}}\, T^{ij}\, \delta g_{ij}^{(0)} =  \pi L \int \dd t \, \alpha\,  \le({1\over \alpha}\mathcal{T}_{\scaleto{\text{3D}}{4pt}}\, \delta \alpha + {1\over \lambda} \mathcal{O}_{\scaleto{\text{3D}} {4pt}} \,\delta \lambda-  \mathcal{J}^t_{\scaleto{\text{3D}}{4pt}}\,\delta \nu \ri)~,
\ee
where the relations among each side of this equation are
\begin{align}\label{eq:ch23}
\mathcal{T}_{\scaleto{\text{3D}}{4pt}}&= - 2\pi L\,\frac{\lambda}{\alpha^2}\,(T_{tt}+\nu^2 T_{zz}-2\nu T_{tz})~,\cr
\mathcal{O}_{\scaleto{\text{3D}}{4pt}}&= \frac{ 2\pi L}{\lambda}\,T_{zz}~,\cr
\mathcal{J}^t_{\scaleto{\text{3D}}{4pt}}&= 2\pi  L\,\frac{\lambda}{\alpha^2}\,(T_{tz}-\nu T_{zz})~.
\end{align}
In the 3D theory, the consistent stress tensor $T_{ij}$ \eqref{eq:STvR} arises from a well defined variational principle, for which it is meaningful to apply \eqref{eq:ch23}. Using \eqref{eq:map23} and \eqref{eq:asymptoticvalues} we obtain
\begin{align} \label{eq:t2dconsistent}
\mathcal{T}_{\scaleto{\text{3D}}{4pt}}&= -\frac{L \ell}{8G_3} \frac{ \mathfrak{m}}{\lambda} -\frac{L}{4 G_3\mu \ell}\left( \frac{Q}{\lambda}+\ell\,\frac{\lambda\lambda'\nu'}{2\alpha^3}\right)~, \cr
\mathcal{O}_{\scaleto{\text{3D}}{4pt}} &=\frac{L \ell }{8G_3}\left(\frac{\mathfrak{m}}{\lambda}- \frac{\mathfrak{m}'}{\lambda'} \right)+ \frac{L}{4 G_3\mu\ell}\left(\frac{Q}{\lambda}+ \frac{\ell}{2}\left(\frac{\alpha'\lambda^2 \nu'}{\alpha^4}-\frac{\lambda^2\nu''}{2\alpha^3}\right)\right)~,\cr 
\mathcal{J}^t_{\scaleto{\text{3D}}{4pt}} &= \frac{L}{4 G_3}\frac{Q}{\alpha} +\frac{L}{8 G_3\mu} \left(\frac{m_0}{\alpha} - \frac{\alpha' \lambda\lambda'}{\alpha^4}+ \frac{\lambda'^2}{2\alpha^3}+ \frac{\lambda\lambda''}{2\alpha^3}\right)~.
\end{align}
In terms of these variables, the trace anomaly and Ward identity \eqref{eq:tracean} take the form\footnote{Using \eqref{eq:g0} and \eqref{eq:ch23}, the divergence appearing in \eqref{eq:tracean} translates to 
\begin{align}\nonumber
 D_iT^{it}&=-\frac{1}{\alpha^2\lambda}\left(\partial_t\mathcal{T}-\mathcal{O}\,\partial_t\log\lambda \right),\quad \DD_i T^{iz}=\frac{\nu}{\alpha^2\lambda}\left(\partial_t\mathcal{T}-\mathcal{O}\,\partial_t\log\lambda\right)-\frac{1}{\lambda^3}\left(\partial_t\mathcal{J}^t + \mathcal{J}^t\partial_t\log\alpha\right)~,
\end{align} 
 and the trace is  $T_i^i = \frac{1}{\lambda}\left(\mathcal{T} + \mathcal{O}\right)$.
}
\be
\mathcal{T}_{\scaleto{\text{3D}}{4pt}}+\mathcal{O}_{\scaleto{\text{3D}}{4pt}}=\frac{L\ell}{4G_3}\frac{1}{\alpha}\left(\partial_t\left(\frac{\lambda'}{\alpha}\right) -\frac{1}{4\mu\ell}\partial_t\left(\frac{\lambda^2\nu'}{\alpha^2}\right)\right)~,
\ee
and
 \begin{align}\label{eq:diffexp}
&\partial_t\mathcal{J}^t_{\scaleto{\text{3D}}{4pt}}+\mathcal{J}^t_{\scaleto{\text{3D}}{4pt}}\,\partial_t\log\alpha= \frac{L}{16G_3\mu\alpha}\,\partial_t^2\left(\frac{\lambda\lambda'}{\alpha^2}\right)~,\cr
&\partial_t\mathcal{T}_{\scaleto{\text{3D}}{4pt}}-\mathcal{O}_{\scaleto{\text{3D}}{4pt}}\,\partial_t\log\lambda=- \frac{L}{16G_3\mu }\left[\frac{\alpha}{\lambda\lambda'}\partial_t\left(\nu'\left(\frac{\lambda\lambda'}{\alpha^2}\right)^2\right) \right]~.
 \end{align}

The renormalised boundary action $\tilde{I}_{\scaleto{\text{ren}}{3pt}}$, which can be inferred by integrating \eqref{eq:var23}, reads
 \begin{align}\label{eq:ren11}
\tilde{I}_{\scaleto{\text{ren}}{3pt}}&= -\frac{L\ell}{8G_3}\int \dd t ~\left(\frac{\alpha}{\lambda} \le[m_0 +2{Q\over \mu \ell^2}\ri]+ \frac{\lambda'^2}{\alpha\lambda}+ \frac{2\nu}{\ell} \le[Q + {m_0\over 2 \mu}\ri]\right) + \frac{L}{16 G_3 \mu}\int \dd t \; \frac{\lambda\lambda'\nu'}{\alpha^2}~.
 \end{align} 
Clearly the consistent stress tensor, dimensionally reduced to 2D, does not coincide with the one-point functions in \eqref{eq:cch1}. Albeit the EMD contributions are in perfect agreement, as reported in \cite{Cvetic:2016eiv}, and the $Q$ dependent pieces also agree,  there is an additional term coming from the gravitational Chern-Simons term. The reason of this mismatch is not surprising: The last term in $\tilde{I}_{\scaleto{\text{ren}}{3pt}}$ can be rewritten as
\begin{align}\label{eq:extra1}
 \frac{L}{16 G_3 \mu}\int \dd t\, \frac{\lambda\lambda'\nu'}{\alpha^2} &= - \frac{L}{32G_3 \mu}\int \dd t \, \gamma^{tt}\partial_t(e^{-2\phi})\, \partial_t A_t~,
\end{align}
which is clearly not gauge invariant.  It is as well finite, and methods for holographic renormalisation are not capable to fix finite counterterms unless another principle (or symmetry) is advocated for. From the 2D point of view, ${I}_{\scaleto{\text{rCS}}{4pt}}$ is a gauge invariant action which makes the introduction of \eqref{eq:extra1} somewhat awkward.  The only meaningful observation we can make at this stage is that \eqref{eq:extra1} can be achieved by integrating by parts either the first or second term in \eqref{eq:KKreducedCS} and arranging time derivatives appropriately. This all illustrates that important parts of the anomalies are lost in the 2D theory (\ref{eq:2Daction}), in particular if we don't make reference to the parent theory. Notably our one-point functions (\ref{eq:cch1}) lead to conserved currents (\ref{eq:conservedcurrents}), while in \eqref{eq:diffexp} we still encounter the effects of the gravitational anomaly.

It is instructive to also compare our results with the dimensional reduction of the covariant stress tensor $t_{ij}$ in \eqref{eq:STTMG1}. Even though this choice of stress tensor does not comply with a variational principle, we will simply inquire what the map \eqref{eq:ch23} predicts in 2D. The result is
\begin{align} \nonumber
\mathcal{t}_{\scaleto{\text{3D}}{4pt}}&=-\frac{L \ell}{8G_3} \frac{ \mathfrak{m}}{\lambda} -\frac{L}{4 G_3 \mu \ell}\frac{Q}{\lambda}~,  \quad 
\mathcal{o}_{\scaleto{\text{3D}}{4pt}} = \frac{L \ell }{8G_3}\left(\frac{\mathfrak{m}}{\lambda}- \frac{\mathfrak{m}'}{\lambda'} \right)+ \frac{L}{4 G_3 \mu \ell}\frac{Q}{\lambda}~,\\ \label{eq:charges3d2d}
\mathcal{j}^t_{\scaleto{\text{3D}}{4pt}}  &= \frac{L}{4 G_3}\frac{Q}{\alpha} +\frac{L}{8 G_3 \mu}\left(\frac{\mathfrak{m}}{\alpha} -\frac{\lambda}{2\lambda'\alpha}\,\mathfrak{m'} \right)~,
\end{align}
where we replaced $T_{ij}\to t_{ij}$ in \eqref{eq:ch23}, and used \eqref{eq:STTMG1}, \eqref{eq:g0} and \eqref{eq:g2}.  The trace anomaly and Ward identity (\ref{eq:trace2}) in this case reduce to
\begin{align}
&\mathcal{t}_{\scaleto{\text{3D}}{4pt}}+\mathcal{o}_{\scaleto{\text{3D}}{4pt}}=\frac{L\ell}{4 G_3}\frac{1}{\alpha}\,\partial_t\left(\frac{\lambda'}{\alpha}\right)~,
\end{align}
and
\begin{align}
&\partial_t\mathcal{j}^t_{\scaleto{\text{3D}}{4pt}}+\mathcal{j}^t_{\scaleto{\text{3D}}{4pt}}\,\partial_t\log\alpha= \frac{\lambda^2}{2\alpha}\,\partial_t \left(\frac{2}{\alpha\lambda}\partial_t\left(\frac{\lambda'}{\alpha}\right)\right)~,\quad \partial_t\mathcal{t}_{\scaleto{\text{3D}}{4pt}}-\mathcal{o}_{\scaleto{\text{3D}}{4pt}}\,\partial_t\log\lambda= 0~.
 \end{align}
These relations show that the  gravitational anomaly  enters in the 2D counterparts of $t_{ij}$, i.e. $(\mathcal{t}_{\scaleto{\text{3D}}{4pt}},\mathcal{o}_{\scaleto{\text{3D}}{4pt}}, \mathcal{j}^t_{\scaleto{\text{3D}}{4pt}})$,  only via    the conservation equation for $\mathcal{j}^t_{\scaleto{\text{3D}}{4pt}} $, while other relations remain as in EMD. For this reason,  one can observe that $\mathcal{t}_{\scaleto{\text{3D}}{4pt}}=\mathcal{T}_{\scaleto{\text{UV}}{4pt}}$ and $\mathcal{o}_{\scaleto{\text{3D}}{4pt}}=\mathcal{O}_{\scaleto{\text{UV}}{4pt}}$, which is clear from comparing \eqref{eq:charges3d2d} and \eqref{eq:cch1}, while the currents $\mathcal{j}^t_{\scaleto{\text{3D}}{4pt}}$  and $\mathcal{J}^t_{\scaleto{\text{UV}}{4pt}}$ do not agree. Recall that the covariant stress tensor $t_{ij}$ does not conform with the Wess-Zumino consistency conditions, so it is not surprising to find this disagreement between the 2D analysis resulting in \eqref{eq:cch1}, and  those obtained via dimensional reduction in \eqref{eq:charges3d2d}.

Finally, we have the 3D conserved stress tensor in \eqref{eq:t3}. Because this object is not Lorentz invariant it is not clear how to treat it in the dimensionally  reduced theory. One obstruction is that we cannot use  the map \eqref{eq:ch23}: It assumes the 3D tensor is symmetric, and therefore contradicts the relations in \eqref{eq:456}.

 
\section{Holographic renormalisation: IR perspective}\label{sec:IR}
In addition to the backgrounds considered in Sec.\,\ref{sec:solUV}, the equations of motion $(\ref{eq:MaxDil})$-$(\ref{eq:eomtrace})$ also admit  a  branch of solutions specified by a constant value of the dilaton. We will denote this branch IR fixed points, due to their role in describing the AdS$_2$ geometry of near extremal black holes. 

In this section we will start with a derivation of the IR fixed point solutions, and  then turn on an irrelevant deformation for the dilaton. This deformation drives also the metric away from its locally AdS$_2$ form attained at the fixed point. On this deformed background we will evaluate the appropriate one-point functions using holographic renormalisation. 

\subsection{Background solution}\label{sec:irback}
To construct the IR fixed point solution, we start by setting
\be\label{eq:constantDilaton}
e^{2\phi}=e^{2\phi_0}~,  
\ee
with $\phi_0$ a constant. We will use the subscript `0' to refer to the values of the fields at the IR fixed point. Subtracting two times \eqref{eq:eomtrace} from \eqref{eq:MaxDil} we infer 
\begin{align}
R= -{6\over \ell^2}-{1\over 2} \,e^{-2\phi_0} f^2&~,
\end{align}
which after plugging it back into the gauge field equation of motion \eqref{eq:MaxDil}  implies that the field strength $f$ is constant as well. The values of $f$ and $R$ are then determined by
\begin{align}\label{eq:eomIR}
R_0+{6\over \ell^2}+{1\over 2} \,e^{-2\phi_0} f_0^2&=0~,\cr
\le(-{4\over \ell^2}+e^{-2\phi_0}f_0^2\ri)\le(1+{3\over 2\mu}\,e^{-\phi_0}f_0\ri)&=0~.
\end{align}
There are two classes of solutions to $(\ref{eq:eomIR})$. The first is
\begin{align}\label{eq:branch1}
R_0 = -{8\over \ell^2} ~,\qquad e^{-2\phi_0}f_0^2 ={4\over \ell^2}~,
\end{align}
which is the constant dilaton solution to EMD. The second branch is
\begin{align}
R_0 = -{6\over \ell^2} -{2\mu^2\over 9}~,\qquad e^{-\phi_0}f_0 =-{2\mu \over 3}~.
\end{align}
This configuration would uplift to warped AdS$_3$ solutions in TMG, such as those in \cite{Nutku:1993eb,Gurses:1994bjn,Anninos:2008fx}.
Since the Ricci scalar is negative for real values of the variables, all fixed point solutions are locally AdS$_2$.

The focus for the reminder of this section will be on \eqref{eq:branch1}. 
Working in the same gauge as in \eqref{eq:gaugechoice}, the background AdS$_2$ metric and gauge field are given by
\begin{align}\label{eq:IRsolutions}
\sqrt{-\gamma_0}&= \alpha_{\scaleto{\text{ir}}{5pt}}(t)\,e^{2r/\ell} + \beta_{\scaleto{\text{ir}}{5pt}}(t)\,e^{-2r/\ell}, \cr A_t&= \nu_{\scaleto{\text{ir}}{5pt}}(t) - \ell\,Q\,e^{3\phi_0}\left(\alpha_{\scaleto{\text{ir}}{5pt}} (t)\,e^{2r/\ell}- \beta_{\scaleto{\text{ir}}{5pt}}(t)\,e^{-2r/\ell}\right)~.
\end{align}
The subscript ``{ir}'' here is to distinguish the functions appearing in our IR analysis to those in  \eqref{eq:asymptoticvalues} which are relevant for the UV. Here $Q$ is defined as in \eqref{eq:FEMD} and from \eqref{eq:branch1} we have
\be\label{eq:Qsquared}
Q^2={1\over \ell^2} e^{-4\phi_0}~.
\ee
The functions $\alpha_{\scaleto{\text{ir}}{5pt}}$, and $\nu_{\scaleto{\text{ir}}{5pt}}$ act as sources for the AdS$_2$ metric and gauge field, respectively. $\beta_{\scaleto{\text{ir}}{5pt}}$ 
parametrizes nearly-AdS$_2$ spacetimes: It is induced by large diffeomorphisms that preserve the boundary metric, as we shall see in Sec.\,\ref{sec:schwir}.

Small perturbations around the IR background, will be ignited by a deviation of the dilaton away from its constant value:\footnote{We are adapting the same notation as in \cite{Castro:2018ffi}, and our subsequent derivations follow closely to those there. This is expected since 2D theories of gravity coupled to a dilaton follow universal trends that are present here too \cite{Almheiri:2014cka,Almheiri:2016fws,GrumillerMcNeesSalzerEtAl2017}.}
 \be\label{eq:perturbationsIR1}
 e^{-2\phi}=e^{-2\phi_0}+\mathcal{Y}~,
 \ee
 with $\mathcal Y$ small. As the equations of motion will demand, this perturbation will generate a backreaction of the metric which we parametrise as 
\begin{align}\label{eq:perturbationsIR}
\sqrt{-\gamma}= \sqrt{-\gamma_0}+ \sqrt{-\gamma_1}~.
\end{align}
The response of the gauge field follows automatically from \eqref{eq:FEMD}. Here all fields depend explicitly on time and the radial coordinate, but we suppress it for notational convenience. We will determine the expressions of the perturbations $\sqrt{-\gamma_1}$ and $\mathcal{Y}$ by solving the perturbed EMD equations of motion, which we know leads to a solution of the full 2D theory \eqref{eq:2Daction}.  These linearised equations around the IR fixed point are
\begingroup
\allowdisplaybreaks
\begin{align}\label{eq:gammatt}
\left(\partial_r^2-\frac{4}{\ell^2}\right)\mathcal{Y}=0~,\\  \label{eq:gammarr}
\left(K_0\,\partial_r + D_{t,0}^2-\frac{4}{\ell^2}\right)\mathcal{Y}=0~,\\ \label{eq:offdiagmetric}
 \partial_r\left(\frac{\partial_t \mathcal{Y}}{\sqrt{-\gamma_0}}\right)=0~,\\ \label{eq:dilatonIRpert}
\left(\partial_r^2-\frac{4}{\ell^2}\right)\sqrt{-\gamma_1} + \frac{6}{\ell^2}\,e^{2\phi_0}\,\sqrt{-\gamma_0}\mathcal{Y}=0~.
\end{align}
\endgroup
The subscript `0' for the trace of the extrinsic curvature $K_0$ and the Laplace Beltrami operator $D_{t,0}^2$ indicate again that they are evaluated in the IR geometry with metric $\gamma_0$
\begin{align}
K_0 &\equiv\partial_r\log\sqrt{-\gamma_0}~,\quad \quad 
D_{t,0}^2\equiv \frac{1}{\sqrt{-\gamma_0}}\partial_t\,\left(\sqrt{-\gamma_0}\, \gamma_0^{tt}\,\partial_t\right)~.
\end{align}
Equation \eqref{eq:gammatt} implies
\be\label{eq:YIR}
\mathcal{Y}= \lambda_{\scaleto{\text{ir}}{5pt}}(t)\,e^{2r/\ell}+ \sigma_{\scaleto{\text{ir}}{5pt}}(t)\,e^{-2r/\ell}~,
\ee
with  $\lambda_{\scaleto{\text{ir}}{5pt}}(t)$ the source for our deformation, and $\sigma_{\scaleto{\text{ir}}{5pt}}(t)$ its vev. Also we can infer from this equation that $\mathcal{Y}$ has conformal dimension $\Delta=2$ and is, as already anticipated, an irrelevant deformation moving us slightly away from the IR fixed point. The constraint in \eqref{eq:offdiagmetric} relates $\mathcal{Y}$ to $\gamma_0$ by imposing
\begin{align}\label{eq:coupled}
\beta_{\scaleto{\text{ir}}{5pt}}(t)= \alpha_{\scaleto{\text{ir}}{5pt}}(t)\,\frac{\sigma_{\scaleto{\text{ir}}{5pt}}'(t)}{\lambda_{\scaleto{\text{ir}}{5pt}}'(t)}~.
\end{align}
This relation is a universal feature of nAdS$_2$ holography: It implies that the perturbation moving us away from the IR fixed point is related to the large diffeomorphisms in $\mathrm{AdS}_2$.
Finally combining \eqref{eq:coupled} with \eqref{eq:gammarr} we obtain 
\begin{align} \nonumber
\beta_{\scaleto{\text{ir}}{5pt}}(t)&= 
-\frac{\ell^2}{16\lambda_{\scaleto{\text{ir}}{5pt}}'(t)}\,\alpha_{\scaleto{\text{ir}}{5pt}}(t)\,\partial_t\left(\frac{\mathfrak{q}(t)}{\lambda_{\scaleto{\text{ir}}{5pt}}(t)}\right)~,\\ \label{eq:relationgamma1gamma0}
\sigma_{\scaleto{\text{ir}}{5pt}}(t)&=- \frac{\ell^2}{16\lambda_{\scaleto{\text{ir}}{5pt}}(t)}\,\mathfrak{q}(t)\;,
\end{align}
where we defined
\begin{align}\label{eq:defmathfrakq}
\mathfrak{q}(t)\equiv c_0 + \left(\frac{\lambda_{\scaleto{\text{ir}}{5pt}}'(t)}{\alpha_{\scaleto{\text{ir}}{5pt}}(t)}\right)^2~,
\end{align}
and $c_0$ is an integration constant, independent of the spacetime coordinates. 

Now we examine the dilaton equation of motion $(\ref{eq:dilatonIRpert})$. Its solution determines the form of the metric perturbation. The homogeneous solution is a locally AdS$_2$ metric, equal to the background solution $\sqrt{-\gamma_0}$. The inhomogeneous equation on the other side is solved by
\begin{align}\label{eq:perturbationsIR2}
\left(\sqrt{-\gamma_1}\right)_{\mathrm{inh}}= -\frac{e^{2\phi_0}}{2}\left(\mathcal{Y}\sqrt{-\gamma_0} + \frac{\ell^2}{2}\,\partial_t\left(\frac{\lambda_{\scaleto{\text{ir}}{5pt}}'(t)}{\alpha_{\scaleto{\text{ir}}{5pt}}(t)}\right)\right)~.
\end{align}

\subsection{Renormalised observables}
We will now perform holographic renormalisation around the perturbed IR background. 
Our starting point is familiar: As in Sec.\,\ref{sec:HRuv} we will build a 2D action, such that for the deformed IR background, the variation
\begin{align}\label{eq:varEMDIR}
\delta I_{\scaleto{\text{2D}}{4pt}}^{\scaleto{\text{IR}}{4pt}}= \int \dd t \left( \pi^{tt}\,\delta\gamma_{tt} + \pi_\phi \,\delta \phi + \pi^t  \,\delta A_t\right)~,
\end{align}
leads to a well defined variational principle. Here the lower case, relative to the upper case in \eqref{eq:varEMDUV}, is to emphasis that the values of the canonical momenta will depend on our boundary conditions, and this affects the renormalisation of the action. 

The boundary conditions on the metric and dilaton follow from \eqref{eq:IRsolutions} and \eqref{eq:YIR}, 
\be\label{eq:bndrycondIR}
\delta\gamma_{tt}= -2\alpha_{\scaleto{\text{ir}}{5pt}}\,e^{4r/\ell} \delta\alpha_{\scaleto{\text{ir}}{5pt}}~, \quad \delta e^{-2\phi}= e^{2r/\ell}\delta\lambda_{\scaleto{\text{ir}}{5pt}}~,
\ee
that is, their leading divergences are interpreted as sources, and as $r\to \infty$ we seek for finite responses under those variations. The deviations  of $e^{-2\phi}$ away from its constant value are large in $r$, still we want to treat them as small perturbations around the IR fixed point. As we study the response of the action  we will therefore take
\begin{align}\label{eq:smallpertcond}
e^{2\phi_0}\,|\lambda_{\scaleto{\text{ir}}{5pt}}|\,e^{2r/\ell}~\ll ~1~,
\end{align}
which implies that we will keep only the first order effect of the deformation. 

We have not mentioned the boundary conditions of the gauge field in $(\ref{eq:bndrycondIR})$ because the gauge field exhibits a crucial difference in the IR compared to the UV. This requires a separate discussion on how to treat its boundary conditions. The issue arises because in the AdS$_2$ region $A_t$ is no longer dominated by the source $\nu_{\scaleto{\text{ir}}{5pt}}$, but by  the volume of AdS$_2$. From \eqref{eq:IRsolutions} we have 
\begin{align}\label{eq:gfdivergenceIR}
A_t&= -\ell\,Q\,e^{3\phi_0}\alpha_{\scaleto{\text{ir}}{5pt}}\,e^{2r/\ell}+\nu_{\scaleto{\text{ir}}{5pt}} + O(e^{-2r/\ell})~,
\end{align}
and for the canonical momenta on the AdS$_2$ background\footnote{The $``\pm''$ in \eqref{eq:pi1tmg} is due to selecting a sign as one uses  \eqref{eq:Qsquared} to evaluate the contribution to $\pi^t$ from the rCS. The minus sign corresponds to $Q<0$, while the plus sign corresponds to $Q>0$.}
\be\label{eq:pi1tmg}
\pi^t = {\delta I_{\scaleto{\text{2D}}{4pt}}\over \delta(\partial_r A_t) }=- {L\over 4 G_3} \left(1 \mp {1\over \mu\ell}\right) Q~,
\ee
which reflects that $A_t\sim \sqrt{-\gamma_0}\,\pi^t$ as $r\to \infty$, and the source is being washed away. The problem therefore is the following: We have a space of asymptotic solutions characterised by a charge $Q$ and source $\nu_{\scaleto{\text{ir}}{5pt}}$, which we want to relate to the space of fields; and it is clear that the asymptotic behaviour of $A_t$ and $\pi^t$ does not capture this information. The solution to the dilemma is explained in  \cite{Sen:2008vm,Cvetic:2016eiv, Castro:2018ffi}. Here we will provide a brief summary. 

To fix this issue, it is convenient to first do a canonical transformation 
\begin{align}\label{eq:IbgaugefieldIR}
I_{\rm ren}[\gamma_{tt},\phi,\pi^t]=I_{\scaleto{\text{2D}}{4pt}}^{\scaleto{\text{IR}}{4pt}}-\int\dd t\,\pi^t A_t~,
\end{align}
which is a Legendre transform for the gauge field, and start from the variational problem 
\begin{align}\label{eq:varEMDIRx}
\delta I_{\rm ren}= \int \dd t \left( \pi^{tt}\,\delta\gamma_{tt} + \pi_\phi \,\delta \phi -A_t^{\rm ren}  \,\delta \pi^t\right)~,
\end{align}
Here $A_t^{\rm ren}$ is identified as the conjugate variable to $\pi^t$ and obeys
\begin{align}\label{eq:defAren}
A_t^{\rm ren}= A_t~- ~\frac{I_{\scaleto{\text{2D}}{4pt}}^{\scaleto{\text{IR}}{4pt}}}{\delta \pi^t}~.
\end{align}
Now, as one constructs the renormalised action $I_{\scaleto{\text{2D}}{4pt}}^{\scaleto{\text{IR}}{4pt}}$, one should assure that from \eqref{eq:defAren} we obtain $A_t^{\rm ren}= \nu_{\scaleto{\text{ir}}{5pt}} + O(e^{-2r/\ell})$. Finally, since we want to have a Dirichlet valued problem for all fields, the last step is to do another Legendre transform
\be
\hat I_{\rm ren}= I_{\rm ren}-\int\dd t\,\pi^t A_t^{\rm ren}~,
\ee
and look for finite responses of the effective action, i.e.
\begin{align}\label{eq:varEMDIRy}
\delta \hat I_{\rm ren}= \int \dd t \left( \pi^{tt}\,\delta\gamma_{tt} + \pi_\phi \,\delta \phi+ \pi^t  \,\delta A_t^{\rm ren}\right)~,
\end{align}
where now our boundary conditions for all fields are
\be
\delta\gamma_{tt}= -2\alpha_{\scaleto{\text{ir}}{5pt}}\,e^{4r/\ell} \delta\alpha_{\scaleto{\text{ir}}{5pt}}~, \quad \delta e^{-2\phi}= e^{2r/\ell}\delta\lambda_{\scaleto{\text{ir}}{5pt}}~, \quad \delta A_t^{\rm ren}= \delta  \nu_{\scaleto{\text{ir}}{5pt}}~.
\ee
In the following we will start our construction of counterterms immediately from \eqref{eq:varEMDIRy}, and not build $I_{\rm ren}$  or $I_{\scaleto{\text{2D}}{4pt}}^{\scaleto{\text{IR}}{4pt}}$ explicitly; we refer to \cite{Castro:2018ffi} for those intermediate steps which are easily adapted to the discussion here. 

\subsubsection{One-point functions}

We now turn to building the boundary terms needed to make \eqref{eq:varEMDIRy} a well defined variational problem. It is easier to first  focus on the contributions from EMD, which resembles the analysis in \cite{Castro:2018ffi} adopted to our setup. As in the UV analysis, the Gibbons-Hawking term  guarantees a Dirichlet boundary problem
\begin{align}
I_{\scaleto{\text{GH}}{4pt}}^{\scaleto{\text{IR}}{4pt}}&=\frac{L}{4G_3}\int\dd t\, \sqrt{-\gamma}\,e^{-\phi}\,K~,
\end{align}
and the counterterm that renders the variation finite is
\begin{align}\label{eq:counterIREMD}
I_{\scaleto{\text{d1}}{4pt}}&=  -\frac{L}{4G_3 \ell} \int \dd t\, \sqrt{-\gamma}\,e^{\phi_0}\,\mathcal{Y}+ \frac{L}{8G_3 \ell}\int \dd t\sqrt{-\gamma}\,e^{3\phi_0}\,\mathcal{Y}^2~.
\end{align}
The on- shell variation of the EMD action combined with these terms gives
\begin{align}\nonumber
\delta\left(I_{\scaleto{\text{EMD}}{4pt}}+I_{\scaleto{\text{GH}}{4pt}}^{\scaleto{\text{IR}}{4pt}}+I_{\scaleto{\text{d1}}{4pt}}\right)=&\frac{L}{8G_3}\int \dd t \sqrt{-\gamma}\left(\partial_r e^{-\phi} -\frac{1}{\ell}\,e^{\phi_0}\,\mathcal{Y}\right)\gamma^{tt}\delta\gamma_{tt}\\ 
&-\frac{L}{4G_3}\int \dd t \sqrt{-\gamma}\left(K e^{-\phi}- \frac{2}{\ell}\,e^{-\phi_0}\right)\delta\phi - \frac{L}{4G_3}\int \dd t\, Q\, \delta A_t^{\scaleto{\text{ren}}{3pt}}~. \label{eq:renEMDIR}
\end{align}

For the rCS action, we can start from  \eqref{eq:varfinal}: The combination $I_{\scaleto{\text{rCS}}{4pt}} + I_{\scaleto{\text{c2}}{4pt}}$ fixes Dirichlet boundary conditions. Replacing \eqref{eq:perturbationsIR1}, \eqref{eq:perturbationsIR}, and \eqref{eq:perturbationsIR2} in \eqref{eq:varfinal}, and keeping terms consistent with \eqref{eq:smallpertcond}, gives
\begin{align}\label{eq:varDirichletIRm}
\delta_{\gamma_{tt}}\left(I_{\scaleto{\text{rCS}}{4pt}}+ I_{\scaleto{\text{c2}}{4pt}}\right)&= -\frac{L}{16G_3\mu}\int \dd t \sqrt{-\gamma_0}\, e^{3\phi_0}\,Q\,\partial_r\mathcal{Y}\,\gamma_0^{tt}\delta\gamma_{tt} ~,
\end{align}
and 
\begin{align} \label{eq:varDirichletIRdx}
&\delta_{\phi}\left(I_{\scaleto{\text{rCS}}{4pt}}+ I_{\scaleto{\text{c2}}{4pt}}\right)
= \frac{L}{4G_3\mu}\int \dd t \,e^{3\phi_0} \,Q\left( e^{-2\phi_0}\partial_r\sqrt{-\gamma_0}-\mathcal{Y}\,\partial_r\sqrt{-\gamma_0} -\frac{3}{2}\,\sqrt{-\gamma_0}\,\partial_r\mathcal{Y}\right)\,\delta\phi~.
\end{align}
Contrarily to the on-shell values for the UV, where \eqref{eq:varfinal} led to finite contributions, the above terms are divergent as $r\to \infty$ for the IR values \eqref{eq:IRsolutions} and \eqref{eq:YIR}. Removing these IR divergences will lead to quantitative differences in our one-point functions as will be evident shortly. 
To cure the remaining divergences in \eqref{eq:varDirichletIRm}-\eqref{eq:varDirichletIRdx} we will add the following counterterms
\begin{align}\label{eq:c3}
I_{\scaleto{\text{c3}}{4pt}}= \frac{L}{4G_3\mu\ell}\int \dd t\,  \sqrt{-\gamma}\,e^{3\phi_0}\, Q\,\mathcal{Y} -\frac{3\,L}{8G_3\mu\ell}\int \dd t \, \sqrt{-\gamma}\,e^{5\phi_0}\, Q\,\mathcal{Y}^2~.
\end{align}
It is worth noting that these counterterms are very similar to those in \eqref{eq:counterIREMD} used for EMD. The reason for their similarity is due to the fact that we are working at first order in the perturbations, and these are the allowed combinations of $\cal{Y}$ that could cancel divergences induced by the irrelevant deformation.     

Combining the contributions from \eqref{eq:varDirichletIRm}-\eqref{eq:c3}, plus the contribution from $A_t^{\rm ren}$, we finally have
\begin{align}\nonumber
\delta \left(I_{\scaleto{\text{rCS}}{4pt}}+ I_{\scaleto{\text{c2}}{4pt}}+ I_{\scaleto{\text{c3}}{4pt}}\right)
=&  -\frac{L}{16G_3\mu}\int \dd t \, e^{3\phi_0} \,Q \,\sqrt{-\gamma_0}\,\left(\partial_r\mathcal{Y} - \frac{2}{\ell}\,\mathcal{Y}\right)\,\gamma_0^{tt}\delta\gamma_{tt} \\ \nonumber
&- \frac{L}{4G_3\mu}\int \dd t\, e^{3\phi_0} \,Q \,\left(\mathcal{Y}\,\partial_r\sqrt{-\gamma_0} +\frac{3}{2}\,\sqrt{-\gamma_0}\,\partial_r\mathcal{Y}- 4\, \sqrt{-\gamma_0}\,\mathcal{Y} \right)\,\delta\phi\\ \nonumber
&+ \frac{L}{4G_3\mu}\int \dd t\, e^{\phi_0} \,Q \, \left(\partial_r\sqrt{-\gamma_0}- \frac{2}{\ell}\sqrt{-\gamma_0}- \frac{2}{\ell}\sqrt{-\gamma_1}\right)\,\delta\phi~\\ 
&- \frac{L}{4G_3\mu}\int \dd t \,  e^{2\phi_0}\,Q^2\,\delta A_t^{\scaleto{\text{ren}}{3pt}}~.
\end{align}
 The renormalised one-point functions in the IR are given by\footnote{Note that there is a small difference in the definition of ${\cal O}$ in the UV relative to the IR. This is simply because of the nature of our boundary fall-offs: in the UV we have $\delta \phi= \lambda^{-1}\delta \lambda$, while in the IR $\delta \phi= -{1\over 2} e^{2\phi_0} e^{2r/\ell} \delta \lambda_{\scaleto{\text{ir}}{5pt}}$. }
\begin{align} \label{eq:opfIR1}
\mathcal{T}_{\scaleto{\text{IR}}{4pt}} \equiv \frac{2}{\alpha_{{\scaleto{\text{ir}}{5pt}}}}\lim_{r\rightarrow\infty}{\pi}_{t}^{~t} ~, \quad  \mathcal{O}_{\scaleto{\text{IR}}{4pt}} \equiv-\frac{1}{\alpha_{{\scaleto{\text{ir}}{5pt}}}}\lim_{r\rightarrow\infty}e^{2r/\ell}{\pi}_{\phi}~,\quad 
 \mathcal{J}^{t}_{\scaleto{\text{IR}}{4pt}} \equiv-\frac{1}{\alpha_{{\scaleto{\text{ir}}{5pt}}}}\lim_{r\rightarrow\infty}\pi^{t}~.
\end{align} 
Explicitly, using \eqref{eq:IRsolutions} and \eqref{eq:YIR}  we obtain 
\begin{align}\nonumber
\mathcal{T}_{\scaleto{\text{IR}}{4pt}} &= -\frac{L}{2G_3\ell}\,e^{\phi_0}\left(1-\frac{Q}{\mu}\,e^{2\phi_0}\right)\sigma_{{\scaleto{\text{ir}}{5pt}}}
= \frac{L \ell}{32G_3}\,e^{\phi_0}\left(1\pm \frac{1}{\mu\ell}\right)\,\frac{\mathfrak{q}(t)}{\lambda_{{\scaleto{\text{ir}}{5pt}}}}~,\\ \nonumber
 \mathcal{O}_{\scaleto{\text{IR}}{4pt}} &=-\frac{L}{G_3\ell}\,e^{-\phi_0}\left(1-\frac{Q}{\mu}\,e^{2\phi_0}\right)\frac{\beta_{{\scaleto{\text{ir}}{5pt}}}}{\alpha_{{\scaleto{\text{ir}}{5pt}}}}
 = \frac{L\ell}{16G_3}\, e^{-\phi_0}\left(1\pm \frac{1}{\mu\ell}\right)\,\frac{1}{\lambda_{{\scaleto{\text{ir}}{5pt}}}'}\partial_t\left(\frac{\mathfrak{q}(t)}{\lambda_{{\scaleto{\text{ir}}{5pt}}}}\right)~,\\ \label{eq:opfIR}
 \mathcal{J}^{t}_{\scaleto{\text{IR}}{4pt}} &= \frac{L}{4G_3}\,Q\,\left(1+\frac{Q}{\mu}\, e^{2\phi_0} \right)\frac{1}{\alpha_{{\scaleto{\text{ir}}{5pt}}}} = \frac{L}{4G_3}\,Q\,\left(1\mp\frac{1}{\mu\ell} \right)\frac{1}{\alpha_{{\scaleto{\text{ir}}{5pt}}}} ~.
\end{align} 
In the second equality we used  \eqref{eq:relationgamma1gamma0} which gives the on-shell values of the one-point functions. We also used \eqref{eq:Qsquared} which relates $Q$ to  $e^{-2\phi_0}$ up to a choice of sign for $Q$. 
From here we can deduced that the renormalised on-shell boundary action is
\begin{align}\label{eq:effactionIR}
\hat I_{\rm ren}&=\frac{L \ell}{32G_3}\left(1\pm \frac{1}{\mu\ell}\right)\int \dd t~e^{\phi_0}\,\frac{\alpha_{{\scaleto{\text{ir}}{5pt}}}}{\lambda_{{\scaleto{\text{ir}}{5pt}}}}\left[c_0- \left(\frac{\lambda_{\scaleto{\text{ir}}{5pt}}'}{\alpha_{\scaleto{\text{ir}}{5pt}}}\right)^2\right]+\frac{L}{4G_3}\left(1\mp \frac{1}{\mu \ell}\right)\,\int\dd t\, Q\,\nu_{{\scaleto{\text{ir}}{5pt}}}~,
\end{align}
satisfying 
\begin{align}\label{eq:variationIRrf}
\mathcal{T}_{\scaleto{\text{IR}}{4pt}} &= \frac{\hat{I}_{\mathrm{ren}}}{\delta \alpha_{{\scaleto{\text{ir}}{5pt}}}}~, \qquad 
 \mathcal{O}_{\scaleto{\text{IR}}{4pt}} = \frac{2}{\alpha_{{\scaleto{\text{ir}}{5pt}}}}e^{-2\phi_0} \frac{\delta \hat{I}_{\mathrm{ren}} }{\delta \lambda_{{\scaleto{\text{ir}}{5pt}}}}~,\qquad 
 \mathcal{J}^{t}_{\scaleto{\text{IR}}{4pt}}= \frac{1}{\alpha_{{\scaleto{\text{ir}}{5pt}}}}\frac{\delta \hat{I}_{\mathrm{ren}}}{\delta \nu_{{\scaleto{\text{ir}}{5pt}}}}~.
\end{align}

From \eqref{eq:opfIR}, our one-point functions obey
\begin{align}\label{eq:diffwardIR}
\partial_t \mathcal{T}_{\scaleto{\text{IR}}{4pt}}-\frac{1}{2}\,e^{2\phi_0}\,\lambda_{\scaleto{\text{ir}}{5pt}}'\, \mathcal{O}_{\scaleto{\text{IR}}{4pt}}=0~,\qquad\partial_t\mathcal{J}^t_{\scaleto{\text{IR}}{4pt}} +\mathcal{J}^t_{\scaleto{\text{IR}}{4pt}}\,\partial_t\log\alpha_{\scaleto{\text{ir}}{5pt}}=0~,
\end{align}
and 
\begin{align}\label{eq:tracewardIR}
\mathcal{T}_{\scaleto{\text{IR}}{4pt}}+\frac{1}{2}\,e^{2\phi_0}\lambda_{\scaleto{\text{ir}}{5pt}}\,\mathcal{O}_{\scaleto{\text{IR}}{4pt}}= \frac{L \ell}{16G_3}\,e^{\phi_0}\,\left(1\pm \frac{1}{\mu\ell}\right)\frac{1}{\alpha_{\scaleto{\text{ir}}{5pt}}}\partial_t\left(\frac{\lambda_{\scaleto{\text{ir}}{5pt}}'}{\alpha_{\scaleto{\text{ir}}{5pt}}}\right)~.
\end{align}


\section{Schwarzian effective action}\label{sec:schw}

In this last section we will provide an interpretation of the holographic renormalisation in the UV and IR in terms of the Schwarzian effective action. We will discuss the interpolation between these two fixed points and their role in describing the entropy of the near extremal BTZ black hole.


\subsection{Effective action: IR}\label{sec:schwir}
 We will start by interpreting the results in Sec.\,\ref{sec:IR}, that are relevant for nAdS$_2$ holography. The renormalised on-shell boundary action found in \eqref{eq:effactionIR} takes the form
\begin{align}\label{eq:effactionIR1a}
\hat I_{\rm ren}&=\frac{L \ell}{32G_3}\left(1+ \frac{1}{\mu\ell}\right)\int \dd t~e^{\phi_0}\,\frac{\alpha_{{\scaleto{\text{ir}}{5pt}}}}{\lambda_{{\scaleto{\text{ir}}{5pt}}}}\left[c_0- \left(\frac{\lambda_{\scaleto{\text{ir}}{5pt}}'}{\alpha_{\scaleto{\text{ir}}{5pt}}}\right)^2\right]+\frac{L}{4G_3}\left(1-\frac{1}{\mu \ell}\right)\,\int\dd t\, Q\,\nu_{{\scaleto{\text{ir}}{5pt}}}~,
\end{align}
where we selected $Q<0$; this is the correct choice as we compare to the conventions used for the BTZ black hole in Sec.\,\ref{sec:BTZ}.

To interpret the various pieces in this action, we will first  venture into the asymptotic symmetries relevant to nAdS$_2$ holography. For simplicity we will set $\alpha_{{\scaleto{\text{ir}}{5pt}}}=1$, and start by considering the empty AdS$_2$ background
\be
ds^2= \dd r^2 - e^{4r/\ell}\dd t^2~, \qquad A= -\ell\,Q\,e^{3\phi_0}\,e^{2r/\ell} \dd t~.
\ee
The set of diffeomorphisms that preserve the boundary metric and radial gauge are
\begin{align}\label{Sdiffeo}
t &\to f(t) + {\ell^2\over 8}{ f''(t)  \over  e^{4r/\ell} -{\ell^2\over 16}{f''(t)^2\over f'(t)^2}}~, \cr e^{2r/\ell}&\to {e^{-2r/\ell} \over  f'(t)}\left( e^{4r/\ell} -{\ell^2\over 16}{f''(t)^2\over f'(t)^2}\right)~,
\end{align}
where $f(t)$ labels reparametrizations of the boundary time. The gauge field transforms as well under this diffeomorphism, and to compensate for this, the diffeomorphism needs to be complemented by a  gauge transformation \cite{Hartman:2008dq,Castro:2008ms},
\be
A_\mu \to A_\mu+ \partial_\mu \Lambda ~,\qquad  \Lambda = -{Q\ell\over2}\,e^{3\phi_0}\, \log \le({4 e^{2r/\ell} f'(t)-\ell f''(t) \over 4 e^{2r/\ell} f'(t)+\ell f''(t)} \ri)~,
\ee
 designed to preserve $A_r=0$ and the asymptotic behaviour of the field. The resulting background is
\begin{align}
ds^2&= \dd r^2 - \le( e^{2r/\ell}+{\ell^2\over 8}\{f(t),t\}  e^{-2r/\ell} \ri)^2\dd t^2~, \cr A&= -\ell\,Q\,e^{3\phi_0}\le( e^{2r/\ell}-{\ell^2\over 8}\{f(t),t\}  e^{-2r/\ell} \ri)\dd t~,
\end{align}
which clearly fits \eqref{eq:IRsolutions} with
\be\label{eq:betaschw}
\beta_{{\scaleto{\text{ir}}{5pt}}} = {\ell^2\over 8}\{f(t),t\}~, \qquad  \{f(t),t\}=\le({f''\over f'}\ri)' -{1\over 2}\le({f''\over f'}\ri)^2~.
\ee
This makes explicit that $\beta_{{\scaleto{\text{ir}}{5pt}}}$ is induced by a large diffeomorphism, and its value is given by the Schwarzian derivative of $f(t)$.   It is also instructive to revisit \eqref{eq:relationgamma1gamma0}: Taking a derivative to remove $c_0$ from the first equation implies
\be
\ell^2 \lambda_{{\scaleto{\text{ir}}{5pt}}}'''+8 \lambda_{{\scaleto{\text{ir}}{5pt}}} \beta_{{\scaleto{\text{ir}}{5pt}}}'+16 \beta_{{\scaleto{\text{ir}}{5pt}}}  \lambda_{{\scaleto{\text{ir}}{5pt}}}'=0~,
\ee 
which via \eqref{eq:betaschw} becomes
\be\label{eq:label1}
\le({1\over f'}\le({(f' \lambda_{{\scaleto{\text{ir}}{5pt}}})'\over f'}\ri)'\ri)'=0~.
\ee
As expected from all other instances of nAdS$_2$ holography,  the dynamics of the irrelevant deformation ignited by $\cal Y$ is  related to the reparametrizations of boundary time. 

The dynamics in \eqref{eq:label1} is elegantly encoded in $\hat I_{\rm ren}$, which can be seen as follows. Solving \eqref{eq:relationgamma1gamma0} for $c_0$ and substituting it into
 \eqref{eq:effactionIR1a} leads to
\begin{align}\label{eq:effactionIR1}
\hat I_{\rm ren}&=\frac{L \ell}{32G_3}\left(1+ \frac{1}{\mu\ell}\right)\int \dd t\,e^{\phi_0}\,\le({c_0\over\lambda_{\scaleto{\text{ir}}{5pt}}}- \frac{ \lambda_{\scaleto{\text{ir}}{5pt}}'^2}{\lambda_{\scaleto{\text{ir}}{5pt}}}\ri) \cr
 &=\frac{L \ell}{32G_3}\left(1+ \frac{1}{\mu\ell}\right)\int \dd t\,e^{\phi_0} \, \le(  {16\over \ell^2}\lambda_{\scaleto{\text{ir}}{5pt}} \beta_{\scaleto{\text{ir}}{5pt}} +2\lambda_{\scaleto{\text{ir}}{5pt}}'' -2{\lambda_{\scaleto{\text{ir}}{5pt}}'^2\over \lambda_{\scaleto{\text{ir}}{5pt}}}\ri) \cr
 &=\frac{L \ell}{16G_3}\left(1+ \frac{1}{\mu\ell}\right)\int \dd t\,e^{\phi_0}\, \le(\lambda_{\scaleto{\text{ir}}{5pt}}\, \{f(t),t\} -{\lambda_{\scaleto{\text{ir}}{5pt}}'^2\over \lambda_{\scaleto{\text{ir}}{5pt}}}\ri) ~.
 \end{align}
Recall that we have $ \alpha_{{\scaleto{\text{ir}}{5pt}}}=1$, and we have also ignored $ \nu_{{\scaleto{\text{ir}}{5pt}}}$ since it is not important for this portion. In the second line we used  \eqref{eq:relationgamma1gamma0},  and in the last equality we ignored total derivatives and used \eqref{eq:betaschw}. The variation of this last term with respect to $f(t)$ leads to \eqref{eq:label1}: This is one of the renown features of nAdS$_2$ holography --the Schwarzian action captures the bulk dynamics of the irrelevant deformation \cite{Maldacena:2016upp}.

Finally, it is useful to cast the coupling in \eqref{eq:effactionIR1} in terms of the CFT$_2$ central charges in Sec.\,\ref{sec:hr3d}; this gives
\be
\frac{\ell}{16G_3}\left(1+ \frac{1}{\mu\ell}\right)= {c_L\over 24}~,
\ee
a clear indication that the left moving sector of the CFT$_2$ is controlling the nCFT$_1$. 


\subsection{Effective action: UV}
In (\ref{eq:uvrenac}) we obtained the renormalised on-shell boundary action:
\begin{align}\label{eq:uvact123}
I_{\scaleto{\text{2D}}{4pt}}^{\scaleto{\text{UV}}{4pt}}&=-\frac{L}{8G_3}\ell\int \dd t \,\left(\frac{\alpha}{\lambda} \le[m_0 +2{Q\over \mu \ell^2}\ri]+ \frac{\lambda'^2}{\alpha\lambda}+ \frac{2\nu}{\ell} \le[Q + {m_0\over 2 \mu}\ri]\right)~.
\end{align}
To make the Schwarzian action manifest in the UV we will take a slightly different route relative to the IR. In our derivations in the prior subsection we started by considering the set of diffeomorphisms (plus gauge transformations) that preserve the AdS$_2$ background; this allowed us to relate the irrelevant deformation $\lambda_{{\scaleto{\text{ir}}{5pt}}}$ to the Schwarzian derivative. For the UV background in \eqref{eq:asymptoticvalues} we will instead inquire how the asymptotic background responds to Weyl transformations of the boundary fields, which is the strategy in \cite{Cvetic:2016eiv}. 

A Weyl rescaling of the boundary parameters \eqref{eq:asymptoticvalues} corresponds to bulk diffeormorphisms that preserve the Fefferman-Graham gauge, i.e. a PBH transformation in the nomenclature of \cite{Cvetic:2016eiv}. The response of the sources under this transformation is the expected one: We would have
\be
\alpha ~\to ~\alpha\, e^{\sigma(t)} ~, \quad \lambda ~\to ~\lambda\, e^{\sigma(t)}~, \quad \nu ~\to ~\nu~,
\ee
where $\sigma(t)$ is an arbitrary function that rescales the boundary metric. 
In order to make explicit how to interpret this transformation as reparametrizations of the boundary time, we choose
\be
\sigma(t) = \log f'(t)~,
\ee
along the lines of the transformation in \eqref{Sdiffeo}. For the choice $\alpha=\lambda=e^{\sigma(t)}=f'(t)$ the on-shell action, up to a total derivative, is
\begin{align}\label{eq:schwiuv}
 I_{\scaleto{\text{2D}}{4pt}}^{\scaleto{\text{UV}}{4pt}}&=-\frac{L}{8G_3}\ell\int \dd t \,\left(\frac{f''}{f'}\right)^2=\frac{L}{4G_3}\ell\int \dd t \,\{f(t),t\}~.
\end{align}
Here we ignored the terms proportional to $Q$ and $m_0$ in \eqref{eq:uvact123}, since they are unaffected by the Weyl rescaling. Therefore the manifestation of the Schwarzian derivative in this derivation comes as a responses of the system under Weyl transformations of the boundary metric. This is compatible with the CFT$_2$ interpretation of this term, where the coupling of \eqref{eq:schwiuv} in terms of CFT$_2$ central charges in Sec.\,\ref{sec:hr3d} is
\be
\frac{\ell}{4 G_3}= {c_L + c_R\over 12}={c\over 6}~.
\ee
It is important to highlight that the overall coefficient of \eqref{eq:schwiuv}  is distinct from \eqref{eq:effactionIR1}. This is already an indication that the origin of the Schwarzian term in the UV and IR is different. We will elaborate more on this point in the following.

\subsection{Interpolation between UV and IR}

Having done an independent analysis of the UV and IR backgrounds, we now proceed to compare them. In particular, we will illustrate how to obtain the deformed IR backgrounds as a decoupling limit of configurations in the UV.

To start let us consider static (time independent) configurations. In this case the UV backgrounds \eqref{eq:runningD} and \eqref{eq:runningG} become 
\begin{align}\label{eq:constantUV}
e^{-2\phi}&= \lambda^2e^{2r/\ell}\left(1 + {\ell^2\over2\lambda^2} m_0 e^{-2r/\ell} + \frac{\ell^2}{16\lambda^4}\left(\ell^2 m_0^2- 4Q^2\right)e^{-4r/\ell}\right)~,\cr
\sqrt{-\gamma}&= {\alpha \lambda} \le( e^{2r/\ell}-\frac{\ell^2}{16\lambda^4}\left(\ell^2m_0^2- 4Q^2\right)e^{-2r/\ell}\ri) e^{\phi}~,\cr
A_t&= {\ell\,\alpha\over \lambda}\,Q \, e^{2\phi} + \nu~.
\end{align}
To obtain the deformed AdS$_2$ background as a limit of this background, we redefine 
\be\label{eq:limit222}
\ell^2 m_0^2 = 4Q^2+\epsilon^2 ~, \qquad   e^{2r/\ell}\to  {\epsilon\over 4}\,e^{2r/\ell}  ~, \qquad t\to {4\over \epsilon}\, t~,
\ee
and take the limit  $\epsilon\to 0$ while holding $Q$, $\lambda$ and $\alpha$ fixed. The resulting background is the IR solution in Sec.\,\ref{sec:irback}, where we identify
\be\label{eq:mapsourcesXX}
 \alpha_{\rm ir}=e^{\phi_0} \alpha\lambda~, \qquad    \lambda_{\rm ir}={\epsilon\over 4}\,\lambda^2 ~, \qquad   \nu_{\rm ir}={4\over \epsilon}\le(\nu-{\alpha\over \lambda}\ri)~
\ee
and $\ell |Q| = e^{-2\phi_0}$. If we restore time dependence in the UV background, the limit is still given by \eqref{eq:limit222}, and the relation between IR and UV quantities is unchanged. It is instructive to rewrite the relation for $\nu$; we have
\be
\nu={\epsilon\over 4}\le(\nu_{\rm ir}+{\alpha_{\rm ir}\over \lambda_{\rm ir}} e^{-\phi_0}\ri)~.
\ee
This relation indicates that gauge transformations in the UV affect time reparametrizations in the IR.  The effect is that  the gauge anomaly in the UV contributes to the conformal anomaly in the IR.  In particular, for the dimensionally reduced 3D stress tensor in \eqref{eq:t2dconsistent}, replacing  \eqref{eq:limit222}-\eqref{eq:mapsourcesXX} in the renormalised action \eqref{eq:ren11} we recover the Schwarzian effective action in \eqref{eq:effactionIR1}. This illustrates that the conformal piece contained in \eqref{eq:schwiuv} is modified as we flow to the IR.

\subsection{Entropy of 2D black holes}\label{2Dthermo}
 In this last portion we will discuss the ties of the Wald entropy of 2D black holes, and its relation to the Schwarzian action. Comparisons with the entropy of BTZ follow as well. 
 
 For our purposes, a 2D black hole is a static solution with a zero in the metric component $\gamma_{tt}$. Let us start with the UV configurations, where all functions appearing in \eqref{eq:FEMD}-\eqref{eq:defm123} will be considered to be constant, i.e., the solution in \eqref{eq:constantUV}. The existence of a horizon   in \eqref{eq:constantUV} requires $\ell |m_0|\geq 2|Q|$, and its location is 
\be
\gamma_{tt}(r=r_h)=0 ~~\Rightarrow ~~e^{4r_h/\ell}={\ell^2\over 16 \lambda^4}(\ell^2 m_0^2-4Q^2) ~.
\ee
 The temperature we will assign to the black hole is
\be
T_{\scaleto{\text{2D}}{4pt}}^{\scaleto{\text{UV}}{4pt}}= \partial_r \sqrt{-\gamma} |_{r_h}={4\over\ell} \alpha\lambda \, e^{2r_h/\ell}\, e^{\phi_h} ~,
\ee
where the value of the dilaton at the horizon is given by
\be\label{eq:xx1}
e^{-2\phi_h}\equiv e^{-2\phi(r=r_h)} ={\ell\over 2} \le(\ell m_0 + \sqrt{\ell^2 m_0^2-4Q^2}\ri)~.
\ee
 The Wald entropy $S_{\rm Wald}$, which for the 2D action \eqref{eq:2Daction} was derived in \cite{Sahoo:2006vz}, see also  \cite{Park:2006zw,Park:2006gt}, in our notation takes the form
\begin{align}\label{eq:yy1}
S_{\rm Wald}&= \frac{\pi L}{2G_3}\left(e^{-\phi_h}+ \frac{Q}{\mu}\,e^{\phi_h}\right)~.
\end{align}
Note that by substituting \eqref{eq:mapbtzuv2d} in  \eqref{eq:xx1} and \eqref{eq:yy1} reproduces the entropy of the BTZ black hole in \eqref{eq:wald}.

For the IR background, the logic is very similar, the values are just different. We will consider backgrounds \eqref{eq:IRsolutions}-\eqref{eq:defmathfrakq} where all functions are constant. A black hole in this case requires $\beta_{\scaleto{\text{ir}}{5pt}}<0$; the location of the horizon is at 
\be
e^{4r_h/\ell}=-\beta_{\scaleto{\text{ir}}{5pt}} ~,
\ee
which is the zero of $\gamma_{tt}$ in \eqref{eq:IRsolutions} for static configurations. Note that we are adopting $\alpha_{\scaleto{\text{ir}}{5pt}}=1$ to more easily compare with Sec.\,\ref{sec:schwir}. The temperature is
\be
T_{\scaleto{\text{2D}}{4pt}}^{\scaleto{\text{IR}}{4pt}}= \partial_r \sqrt{-\gamma_0} |_{r_h}={4\over\ell} \sqrt{| \beta_{\scaleto{\text{ir}}{5pt}}|}  ~,
\ee
From \eqref{eq:yy1}, the entropy for this background is
\begin{align}
S_{\rm Wald}
&=\frac{\pi L}{2G_3}\left(e^{-\phi_0}\left(1-\frac{1}{\mu\ell}\right)+ \frac{1}{2}\,\mathcal{Y}_h\,e^{\phi_0}\left(1+\frac{1}{\mu\ell}\right)\right)+ O(\mathcal{Y}_h^2)~,
\end{align}
where we used \eqref{eq:perturbationsIR1} and only kept the first correction due to the irrelevant deformation. The value of  ${\mathcal Y}$ at the horizon is
\be
\mathcal{Y}_h= 2\lambda_{\scaleto{\text{ir}}{5pt}} \sqrt{| \beta_{\scaleto{\text{ir}}{5pt}}|} ={\ell \over 2} \lambda_{\scaleto{\text{ir}}{5pt}} T_{\scaleto{\text{2D}}{4pt}}^{\scaleto{\text{IR}}{4pt}}~,
\ee
and so we can write
\be
S_{\rm Wald}=\frac{\pi L}{2G_3}\left(e^{-\phi_0}\left(1-\frac{1}{\mu\ell}\right)+{\ell \over 4} \lambda_{\scaleto{\text{ir}}{5pt}}  \,e^{\phi_0}\left(1+\frac{1}{\mu\ell}\right)T_{\scaleto{\text{2D}}{4pt}}^{\scaleto{\text{IR}}{4pt}}\right)+\ldots~,
\ee
where the dots indicate that this is an expansion around small values of $T_{\scaleto{\text{2D}}{4pt}}^{\scaleto{\text{IR}}{4pt}}$.
From here it is clear that the linear response in the temperature is captured by the IR effective action \eqref{eq:effactionIR1}. In contrast the UV action \eqref{eq:schwiuv}, while it also contains a Schwarzian derivative, does not capture the corrections to the entropy away from extremality.  And finally, using the relations in \eqref{eq:btzirmap1}-\eqref{eq:btzirmap2}, we find perfect agreement with the near-extremal entropy of BTZ given by \eqref{114}. This is all to reinforce that the Schwarzian effective action appearing in nAdS$_2$ holography of the BTZ should be interpreted as follows: It is the response of the left-moving sector of the CFT$_2$ as one deviates away from the zero temperature configuration \cite{Callan:1996dv,Horowitz:1996fn,Larsen:1997ge,Balasubramanian:2003kq,Balasubramanian:2009bg}.

\section*{Acknowledgements}
 We would like to thank Victor Godet, Monica Guica, Daniel Grumiller, Juan Pedraza, and Wei Song for discussions. AC thanks Laboratoire de Physique Th\'eorique de l'Ecole Normale Sup\'erieure for their hospitality while this work was completed. AC work was supported by Nederlandse Organisatie voor Wetenschappelijk Onderzoek (NWO) via a Vidi grant.  This work is supported by the Delta ITP consortium, a program of the NWO that is funded by the Dutch Ministry of Education, Culture and Science (OCW). The work of BM is part of the research programme of the Foundation for Fundamental Research on Matter (FOM), which is financially supported by NWO.

\appendix

\section{BTZ as a 2D black hole}\label{app:btz}

Here we briefly review how to view the BTZ black hole in terms of the 2D variables used in the main portions of the draft. First it is useful to cast  the black hole in the Fefferman-Graham gauge \eqref{eq:FG3d}: Using 
\be
\rho^2= \rho_+^2\cosh^2(\eta/\ell-\eta_0/\ell) -\rho_-^2\sinh^2(\eta/\ell-\eta_0/\ell)~,\qquad e^{2\eta_0/\ell}\equiv {\rho_+^2-\rho_-^2\over 4\ell^2}~,
\ee
in \eqref{eq:BTZ} gives
\begin{align}\label{eq:btzfg}
{ds^2_3} = &{\dd\eta^2}+e^{2\eta/\ell} (-\dd t^2+\ell^2\dd\varphi^2) + \frac{\left(\rho_+ - \rho_-\right)^2}{4\ell^2}\left(\dd t+\ell \dd \varphi\right)^2\cr &+ \frac{\left(\rho_++\rho_-\right)^2}{4\ell^2}\left(\dd t-\ell \dd \varphi\right)^2+\frac{\left(\rho_+^2-\rho_-^2\right)^2}{16\ell^4}e^{-2\eta/\ell}(-\dd t^2+\ell^2\dd\varphi^2)~,
\end{align}
From here we can relate the values of the 2D variables used in Sec.\,\ref{sec:solUV} that lead to the BTZ solution. In relation to \eqref{eq:KKansatz} and \eqref{eq:gaugechoice}, we have  $\eta= r$, $\varphi = z/\ell$, and $L=\ell$. From (\ref{eq:FG3d}) and \eqref{eq:g0} we have
\begin{align}
\alpha=1~,\quad \lambda=1~, \quad \nu=0~,
\end{align}
since $g^{(0)}_{ij}=\eta_{ij}$, and using \eqref{eq:g2} gives
\begin{align}\label{eq:mapbtzuv2d}
\ell^2 m_0= \frac{\rho_+^2 + \rho_-^2}{ \ell^2} = { 8G_3 \, m}~, \cr
\ell^2 Q= -{\rho_+\rho_-\over \ell}=  - 4G_3 \, j~,
\end{align}
where $m$ and $j$ are defined in \eqref{eq:chargesEH}. For $\rho_->0$, as we chose in Sec.\,\ref{sec:BTZ}, we then have $Q<0$.

It is also instructive to map the near horizon geometry of near-extremal BTZ in terms of the variables used in Sec.\,\ref{sec:irback} for the IR deformations. Using the coordinate system in \eqref{eq:btzfg}, the decoupling limit to capture the near horizon is as follows.  We first define the near-extremal black hole as
\be
\rho_\pm= \rho_0 \pm \delta+ O(\delta^2)~.
\ee
 Extremality is at $\delta=0$, and near extremality corresponds to small values of $\delta$. This deviation away from extremality will increase the mass and temperature as described in Sec.\,\ref{sec:btzthermo}. In particular from \eqref{eq:mjT} we have
 \be
 T= {2\delta\over \pi \ell^2} + O(\delta^2)~.
 \ee 
 The dependence on $\delta$ of $\rho_\pm$ is determined by requiring that the angular momentum is fixed for small values of $\delta$. In the coordinate system used here, the horizon is at 
 \be
 e^{2\eta_h/\ell}= e^{2\eta_0/\ell}= {\rho_0\over \ell^2}   \, \delta+ O(\delta^2)
 \ee
 and hence at extremality corresponds to $\eta\to -\infty$. The near horizon region is therefore reached via rescaling our coordinates as
 \be
 \eta \to \eta + \eta_0~, \quad t\to   {\ell t\over \delta}~, \quad \varphi \to \varphi +  {t\over   \delta}
 \ee
and take the limit $\delta\to 0$ in \eqref{eq:btzfg}. The resulting geometry is
\begin{align}\label{eq:nhneBTZ}
ds^2_3\,\underset{\delta\to 0}{\longrightarrow}\, \dd\eta^2 -\gamma_{tt}^{nh}\dd t^2 +r_0^2\left(\dd \varphi + A_t^{nh} \dd t \right)^2+ O(\delta)~,
\end{align}
where
\begin{align}\label{gammtautau}
\gamma_{tt}^{nh}= -(e^{2\rho/\ell}-e^{-2\rho/\ell})^2~,\quad A_t^{nh}=-{1\over r_0}(e^{2\rho/\ell}+e^{-2\rho/\ell})~.
\end{align}
This solution perfectly agrees with the IR fixed point \eqref{eq:IRsolutions}, where we can identify
\be\label{eq:btzirmap1}
\alpha_{\scaleto{\text{ir}}{5pt}}=-\beta_{\scaleto{\text{ir}}{5pt}}=1~, \quad e^{-2\phi_0}= {r_0^2\over \ell^2} ~, \quad Q=-{r_0^2\over \ell^3}~.
\ee
 The first correction in $\delta$ can also be matched with the irrelevant deformation \eqref{eq:perturbationsIR1}. We find
 \be\label{eq:btzirmap2}
 \mathcal{Y}_{nh}=r_0 \delta\, (e^{2\rho/\ell}+e^{-2\rho/\ell})~.
 \ee
And from here we identify $\lambda_{\scaleto{\text{ir}}{5pt}}=\sigma_{\scaleto{\text{ir}}{5pt}}= r_0 \delta$.

\bibliographystyle{JHEP-2}
\bibliography{references}

\end{document}